\documentclass[aps,pra,twocolumn,showpacs,preprintnumbers,amsmath,amssymb]{revtex4-1}
\usepackage{graphicx}
\usepackage[usenames,dvipsnames]{xcolor}
\usepackage{hyperref}
\usepackage{cleveref}
\usepackage{makecell}
\usepackage{float}
\usepackage{lineno}
\usepackage[T1]{fontenc}
\usepackage[caption=false]{subfig}
\newcommand{\ket}[1]{\ensuremath{| #1 \rangle}}

\usepackage{amssymb,amsmath} 
\usepackage{amsfonts}
\usepackage{color}
\usepackage[utf8]{inputenc}
\usepackage{lmodern}
\usepackage{epsfig}
\usepackage{ulem}
\usepackage{blindtext}
\begin{document}
\title{Quantum phases in the interacting generalized Su--Schrieffer--Heeger model}

\author{Jing-Hua Niu$^1$}
\author{Jia-Lin Liu$^1$}
\author{Ke Wang$^2$}
\author{Shan-Wen Tsai$^3$}
\author{Jin Zhang$^1$}
\email{jzhang91@cqu.edu.cn}
\affiliation{$^1$ Department of Physics and Chongqing Key Laboratory for Strongly Coupled Physics, Chongqing University, Chongqing 401331, People's Republic of China}
\affiliation{$^2$ Department of Physics and James Franck Institute, University of Chicago, Chicago, Illinois 60637, USA}
\affiliation{$^3$ Department of Physics and Astronomy, University of California, Riverside, California 92521, USA}
\definecolor{burnt}{cmyk}{0.2,0.8,1,0}
\def\lt{\lambda ^t}
\def\note{note}
\def\beq{\begin{equation}}
\def\enq{\end{equation}}

\begin{abstract}

We investigate the quantum phases of a half-filled generalized interacting Su-Schrieffer-Heeger model with intracell, nearest-neighbor, and next-nearest-neighbor intercell hoppings, together with an on-site inter-sublattice interaction. In the noninteracting limit, the model hosts one topologically trivial phase and two symmetry-protected topological (SPT) phases, distinguished under periodic boundary conditions by different winding numbers and under open boundary conditions by two-fold and four-fold entanglement-spectrum degeneracies, respectively. When interactions are introduced, these free-fermion SPT phases evolve into distinct interacting topological phases that retain characteristic signatures such as entanglement-spectrum degeneracy structures, boundary modes, and nonzero string order parameters. For strong repulsive interactions, a symmetry-breaking phase with unequal but spatially uniform sublattice densities appears between the trivial and topological regimes. For strong attractive interactions, period-2 and period-4 charge-density-wave phases emerge from particle clustering. At intermediate attractive interactions, the competition between interaction-induced localization and hopping-induced delocalization gives rise to a Luttinger liquid phase, a paired Luttinger liquid phase, and a gapless symmetry-protected topological (gSPT) phase. The gSPT phase is characterized by a gapless charge mode together with symmetry-protected current-carrying edge states. We further characterize the gapless phases and the associated quantum phase transitions through central charges and critical exponents.

\end{abstract}


\maketitle

\section{Introduction}\label{sec:introduction}

Understanding how topology intertwines with quantum criticality is a central problem in modern condensed matter physics \cite{OstrovskyInteractionInduced2010PRL,GoswamiQuantum2011PRL,AltlandQuantum2014PRL,DivicAnyon2025PNASU,ZhouTopological2025CP,KirschbaumEmergent2026NP}. While the Landau–Ginzburg–Wilson framework successfully classifies phases and phase transitions associated with spontaneous symmetry breaking, it fails to capture quantum phases whose defining properties are inherently topological. Symmetry-protected topological (SPT) phases exemplify this distinction: they are short-range entangled in the bulk, lack local order parameters, yet host robust boundary phenomena protected by symmetries \cite{SenthilSymmetryProtected2015ARCMP}. Over the past decade, SPT phases in one dimension have been systematically classified \cite{SchnyderClassification2008PRB,KitaevPeriodic2009ACP,FidkowskiEffects2010PRB,TurnerTopological2011PRB,FidkowskiTopological2011PRB,ChenClassification2011PRB,Mondragon-ShemTopological2014PRL,MorimotoBreakdown2015PRB} and experimentally explored \cite{AtalaDirect2013NP,LederRealspace2016NC,MeierObservation2016NC,XieTopological2019nQI,DeLeseleucObservation2019S,LiDirect2023LSA,KlaassenRealization2025NC}, establishing a firm foundation for interacting topological matter.

More recently, increasing attention has been devoted to the role of topology in gapless systems and at quantum critical points. It is now understood that symmetry and microscopic filling constraints can impose fundamental restrictions on the low-energy theory, as first revealed by the Lieb–Schultz–Mattis (LSM) theorem and its extensions~\cite{LiebTwo1961AP,AffleckQuantum1989JPCM,YamanakaNonperturbative1997PRL,OshikawaCommensurability2000PRL,HastingsLiebSchultzMattis2004PRB}. From a modern perspective, these constraints can be interpreted in terms of ’t Hooft anomalies or projective symmetry actions in the effective field theory, which preclude a trivially gapped and symmetric realization~\cite{ChengLiebSchultzMattis2023SP,Elsehooft2025ARCMP}. In general, such systems must either remain gapless or spontaneously break symmetry. Beyond these anomaly-enforced scenarios, gapless phases can also acquire nontrivial topological structure through interaction-driven mechanisms. In particular, multicomponent one-dimensional systems may undergo selective gapping of certain sectors while others remain critical, giving rise to symmetry-protected boundary degrees of freedom embedded within a gapless bulk~\cite{KestnerPrediction2011PRB,ChengMajorana2011PRB,IeminiLocalized2015PRL,MontorsiSymmetryprotected2017PRB,RuhmanTopological2017PRB,Keselmanonedimensional2018PRB,JiangSymmetry2018SB,ThorngrenIntrinsically2021PRB}. A complementary route is provided by decorated domain-wall constructions, which generate topological critical states by twisting otherwise trivial gapless phases~\cite{ScaffidiGapless2017PRX,ParkerTopological2018PRB,LiDecorated2024SP,LiIntrinsically2025SP,WenClassification2025PRB}. In addition, quantum critical points separating distinct SPT or symmetry-breaking phases may themselves exhibit symmetry-enriched structure, where critical bulk modes coexist with protected boundary signatures \cite{VerresenGapless2021PRX}. These developments highlight that topology in gapless systems can emerge from distinct microscopic mechanisms, motivating the search for concrete lattice realizations in paradigmatic models.

The Su–Schrieffer–Heeger (SSH) model provides a paradigmatic realization of a one-dimensional SPT phase. In its minimal form, topology is governed by dimerization: when the intercell hopping exceeds the intracell hopping, the system enters a topological phase with nontrivial bulk winding number and protected boundary modes, whereas the opposite dimerization yields a trivial insulator. Originally introduced to describe polyacetylene, the SSH model has become a canonical framework for studying bulk–boundary correspondence and many-body polarization in one dimension~\cite{SuSolitons1979PRL,HeegerSolitons1988RMP,RestaQuantumMechanical1998PRL}. Extensions incorporating longer-range hopping enrich the band topology, enabling multiple SPT phases distinguished by distinct winding numbers within a single lattice model and giving rise to topological transitions beyond the simplest dimerization-driven scenario~\cite{LiTopological2014PRB,HsuTopological2020PRB,AhmadiTopological2020PRB,QiTopological2021PRR,DiasLongrange2022PRB,ZhengEngineering2022PRA,PellerinWaveFunction2024PRL,JoshiAdiabatic2025PRB}.

Interactions introduce additional competing tendencies. In short-range SSH chains, repulsive interactions can renormalize boundary excitations, compete with dimerization, and drive correlation-induced topological transitions~\cite{ManmanaTopological2012PRB,YoshidaCharacterization2014PRL,YahyaviVariational2018JPCM,KunoPhase2019PRB,NersesyanPhase2020PRB,YuTopological2020PRB,ZhouExploring2023PRB}. To characterize topology beyond single-particle invariants, a range of many-body diagnostics has been developed, including excess boundary charge and many-body polarization under twisted boundary conditions, Green’s-function-based expressions of many-body topological invariants and real-space topological markers, as well as degeneracies of entanglement spectrum (ES)  and entropy scaling~\cite{GurarieSingleparticle2011PRB,YoshidaCharacterization2014PRL,ZhangTwoleg2017PRA,ZegarraCorroborating2019PRB,MeloTopological2023PRB,ZhouExploring2023PRB,DiSalvoTopological2024PRB,WangGroundstate2025PRB}. Complementary field-theoretical approaches based on bosonization and continuum mappings clarify how interactions reorganize low-energy sectors and generate competing ordered or gapless phases in interacting SSH chains \cite{JinBosonization2023PRB}. Related correlated insulating and metallic regimes at commensurate fillings away from half filling have also been explored~\cite{MikhailSuSchriefferHeegerHubbard2024PRB,WangGroundstate2025PRB}. Building on these developments, attention has shifted to the interplay between interactions and longer-range hopping in extended SSH chains, where the competition among multiple topological phases and symmetry-breaking orders gives rise to a rich landscape of quantum phases and critical phenomena, including transitions between trivial, topological, and ordered states~\cite{WangQuantum2023PRB,ZhouInteractioninduced2025PRB,MohamadiEmergence2025PRB}. Existing studies, however, have largely focused on restricted hopping ratios and predominantly repulsive interactions near selected critical points. A systematic exploration that simultaneously varies hopping amplitudes and interaction strength, covering both repulsive and attractive regimes and mapping out the resulting global phase diagram, remains absent. Addressing this gap constitutes the central objective of the present work.

In this work, we present a comprehensive numerical study of a half-filled generalized interacting SSH model with intracell, nearest-neighbor, and next-nearest-neighbor intercell hoppings and an on-site inter-sublattice interaction. We map out its global phase diagram and uncover multiple symmetry-breaking, gapless, and topological phases, together with the associated quantum phase transitions and critical properties. In the noninteracting limit, the model hosts one trivial phase and two free-fermion SPT phases distinguished by different winding numbers. Upon introducing interactions, these evolve into two distinct interacting topological phases, denoted SPT$_1$ and SPT$_2$, which retain characteristic signatures such as ES degeneracy structures, boundary modes, and nonzero string order parameters. For sufficiently strong repulsive interactions, we identify a sublattice-polarized (SP) phase with unequal but spatially uniform densities on the two sublattices, which intervenes between the trivial and topological regimes in part of the phase diagram. In the strongly attractive regime, particle clustering stabilizes two charge-density-wave phases with enlarged unit cells, denoted CDW$_2$ and CDW$_4$. Beyond these insulating phases, an intermediate attractive interaction regime hosts several gapless correlated phases, including a Luttinger liquid (LL) phase, a paired Luttinger liquid (pLL) phase, and a gapless symmetry-protected topological (gSPT) phase. The gSPT phase features a gapless charge sector coexisting with symmetry-protected current-carrying edge states. We characterize the gapless phases and the associated quantum phase transitions by extracting central charges and critical exponents, thereby identifying the corresponding universality classes.

The remainder of this paper is organized as follows. In Sec.~\ref{sec:model}, we introduce the model, discuss its symmetries and noninteracting-limit properties, and summarize the numerical methods and diagnostics. In Sec.~\ref{subsec:phasediagram}, we present the global phase diagrams for several representative parameter slices and discuss the physical origin of the various quantum phases. In Sec.~\ref{subsec:gappedphase}, we analyze the gapped phases, including the interacting topological and symmetry-breaking regimes. In Sec.~\ref{subsec:gaplessphase}, we characterize the gapless phases, including the conventional Luttinger-liquid phases and the gapless topological phase, with emphasis on their critical and topological properties. In Sec.~\ref{subsec:phasetransition}, we analyze the quantum phase transitions out of the symmetry-breaking phases and the gapless phases. Finally, we summarize our results and discuss their broader implications in Sec.~\ref{sec:conclusion}.

\section{Model and Methods}\label{sec:model}

\subsection{Model Hamiltonian} \label{subsec:extsshmodel}

We study a generalized spinless SSH model on a one-dimensional lattice with two sublattices \(A\) and \(B\) per unit cell and \(L\) unit cells in total. The Hamiltonian reads
\begin{eqnarray}\label{eq:extsshham}
\nonumber \hat{H} &=& t_0 \sum_{i=1}^{L} \left( \hat{c}^\dagger_{A,i} \hat{c}_{B,i} + \text{h.c.} \right) + t_1 \sum_{i=2}^{L} \left( \hat{c}^\dagger_{A,i} \hat{c}_{B,i-1} + \text{h.c.}\right) \\
&& + t_2 \sum_{i=3}^{L} \left(\hat{c}^\dagger_{A,i} \hat{c}_{B,i-2} + \text{h.c.}\right) + u \sum_{i=1}^{L} \hat{n}_{A,i} \hat{n}_{B,i},
\end{eqnarray}
where \(t_0\), \(t_1\), and \(t_2\) denote the intra-cell, nearest-neighbor (NN) inter-cell, and next-nearest-neighbor (NNN) inter-cell hopping amplitudes, respectively, and \(u\) is the intra-cell density-density interaction strength. Throughout this work, we set \(t_0 = 1\) to fix the energy scale and consider positive hopping amplitudes, both repulsive and attractive interactions, and half filling with \(L\) fermions in the system.

In the noninteracting limit \(u=0\), the Hamiltonian possesses chiral symmetry. Under a fermion parity transformation acting on one sublattice only, e.g., \(\hat{c}_{B,i} \to -\hat{c}_{B,i}\), the Hamiltonian changes sign. The corresponding chiral symmetry operator is \(\hat{\Gamma} = \exp(\mathrm{i}\pi \sum_j \hat{n}_{B,j})\), which satisfies \(\hat{\Gamma} \hat{H} \hat{\Gamma}^{-1} = -\hat{H}\), or equivalently \(\{\hat{H}, \hat{\Gamma}\}=0\). As a consequence, each positive-energy eigenstate has a negative-energy partner, and the spectrum consists of two bands symmetric about zero energy, separated by a finite gap except at band-touching points. With periodic boundary conditions (PBCs), the noninteracting Hamiltonian can be written in momentum space as
\begin{eqnarray}
\label{eq:kham}
\nonumber \hat{H}_{u=0} &=& \sum_k \Psi_k^\dagger h(k) \Psi_k,
\end{eqnarray}
where \(\Psi_k^\dagger = (\hat{c}_{A,k}^{\dagger}, \hat{c}_{B,k}^{\dagger})\), and
\begin{eqnarray}
h(k) = \mathbf{d}(k)\cdot \mathbf{\sigma} = d_x(k)\sigma_x + d_y(k) \sigma_y,
\end{eqnarray}
with \(d_x(k) = t_0 + t_1 \cos k + t_2 \cos 2k\) and \(d_y(k) = t_1 \sin k + t_2 \sin 2k\). The chiral symmetry is manifested as \(\sigma_z h(k) \sigma_z = -h(k)\). The topological invariant is given by the winding number of the vector \(\mathbf{d}(k)\) as \(k\) traverses the Brillouin zone. Due to the \(\cos 2k\) and \(\sin 2k\) terms, the maximal winding number is two, leading to three distinct insulating phases characterized by \(w=0,1,2\). Within the tenfold-way classification, the noninteracting model belongs to the BDI symmetry class with an integer (\(\mathbb{Z}\)) topological classification \cite{SchnyderClassification2008PRB,KitaevPeriodic2009ACP}. In principle, longer-range hoppings allow arbitrarily large winding numbers, although in the present model with hoppings up to NNN range the winding number is restricted to \(w=0,1,2\).

\begin{figure}[t]
\centering 
\includegraphics[width=1\linewidth]{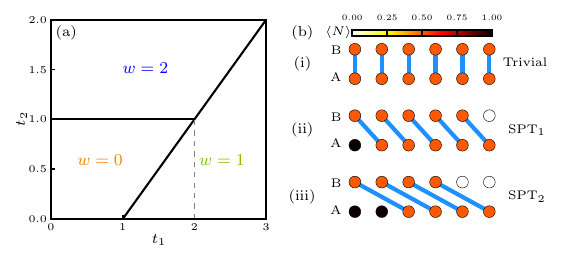}
\caption{Phase diagram in the noninteracting limit with $t_0=1$.
(a) Three topologically distinct band insulators with winding numbers $w=0,1,2$ are separated by the phase boundaries $t_1=t_2+1$ for $t_2>0$ and $t_2=1$ for $0<t_1<2$. 
(b) Schematic bonding structures and symmetry-protected edge modes of the three phases in (a) under OBCs. The phases with $w=0,1,2$ have 0, 1, and 2 edge modes, respectively, and are therefore labeled as trivial, SPT$_1$, and SPT$_2$.} 
\label{fig:noninteracting}
\end{figure}

These insulating phases cannot be adiabatically connected without closing the bulk gap. Diagonalizing \(h(k)\) yields two energy bands,
\begin{equation}
\begin{aligned}
E_{\pm}(k)=\pm \Big[&
t_0^2+t_1^2+t_2^2
+2t_1(t_0+t_2)\cos k \\
&+2t_0t_2\cos 2k
\Big]^{1/2}.
\end{aligned}
\end{equation}
The band-touching points determine the phase diagram of the noninteracting model. As shown in Fig.~\ref{fig:noninteracting}(a) for \(u=0\) and \(t_0=1\), the phase boundary \(t_2 = 1\) with \(t_1 \in (0,2)\) separates the \(w=0\) and \(w=2\) phases. The band touching occurs at \(k_0 = \pm \cos^{-1}(t_1/2t_0)\), which varies continuously with \(t_1/t_0\), giving rise to incommensurate correlations. The other phase boundary is given by \(t_2 = t_1 - 1\). For \(1 < t_1 < 2\), it separates the \(w=0\) and \(w=1\) phases, while for \(t_1 > 2\) it separates the \(w=1\) and \(w=2\) phases. In these cases, the band touching occurs at \(k_0=\pi\). Near the band-touching points, the dispersion is linear. The corresponding low-energy theory is described by conformal field theory with central charge \(c=1\) for \(k_0=\pi\) and \(c=2\) for \(k_0=\pm \cos^{-1}(t_1/2t_0)\), reflecting the presence of one and two independent Dirac fermion modes, respectively. At the intersection point \(t_0 = t_2 = t_1/2\), the dispersion becomes quadratic, defining a Lifshitz critical point with dynamical exponent \(z=2\). This Lifshitz criticality is unstable to interactions and flows to a conformal fixed point upon turning on \(u\) \cite{WangQuantum2023PRB}.

Under open boundary conditions (OBCs), the topology is manifested through boundary modes. When one of the hopping amplitudes dominates, the ground state develops bond order on the corresponding links. For dominant \(t_0\), fermions form intra-cell dimers, yielding a topologically trivial insulating phase. In contrast, dominant \(t_1\) (\(t_2\)) produces one (two) exponentially localized boundary modes at each edge, corresponding to the \(w=1\) (\(w=2\)) phase [see Fig.~\ref{fig:noninteracting}(b)]. These boundary structures can also be diagnosed by the excess boundary charge, defined as the integrated density deviation near one edge relative to the bulk density. In the \(w=1\) phase, the boundary mode carries quantized excess charge \(\pm 1/2\). In the \(w=2\) phase, the edge manifold includes sectors with excess boundary charge \(0\) and \(\pm 1\), reflecting the richer structure associated with two boundary modes per edge.

When interactions are introduced, chiral symmetry is broken, while particle-number conservation and inversion symmetry remain. In one-dimensional chiral-symmetric insulators with conserved \(U(1)\) charge, interactions reduce the free-fermion \(\mathbb{Z}\) classification to \(\mathbb{Z}_4\) \cite{MorimotoBreakdown2015PRB}. This implies that the interacting descendant of the \(w=2\) phase remains topologically distinct from both the trivial phase and the \(w=1\) phase. Although the noninteracting winding number is no longer well defined away from \(u=0\), the interacting phases inherit robust boundary structure from their free-fermion counterparts and will be characterized below by boundary properties, string order parameters, and entanglement diagnostics.

Strong interactions qualitatively modify this picture. For sufficiently strong repulsive interactions \(u>0\), double occupancy within each unit cell is suppressed, leaving the local states \(\ket{n_{A,i}\,n_{B,i}}=\ket{00}\), \(\ket{10}\), and \(\ket{01}\), where the first (second) entry denotes the occupation on sublattice \(A\) (\(B\)). Since inter-cell hopping disfavors configurations with simultaneous occupation of both sublattices in neighboring cells, the effective interactions promote unequal sublattice occupation and stabilize an inversion-symmetry-breaking sublattice-polarized (SP) phase. For strong attractive interactions \(u<0\), the doubly occupied state \(\ket{11}\) is energetically favored. However, inter-cell hopping penalizes configurations in which doubly occupied cells appear on adjacent sites, generating an effective repulsion between bound pairs. This favors spatial separation of doublons and leads to charge-density-wave (CDW) phases. Depending on which inter-cell hopping dominates, different ordering wave vectors are stabilized, resulting in CDW\(_2\) and CDW\(_4\) phases with density modulation periods 2 and 4, respectively.

Between the CDW phases and the gapped SPT phases, the competition between hopping and interactions produces extended gapless regimes. When the commensurate density-wave operators responsible for CDW order become irrelevant, the charge mode remains ungapped and the system forms a Luttinger-liquid (LL) phase. Attractive interactions can instead favor pairing-type processes that gap the relative (sublattice) sector while leaving the total charge sector gapless, yielding a paired Luttinger-liquid (pLL) regime. When the relative sector is instead gapped by a symmetry-preserving topological mass, the bulk remains gapless in the charge channel but supports protected boundary structure, realizing a gapless SPT (gSPT) phase. The microscopic structures and low-energy properties of these phases are analyzed in detail in the following sections.

\subsection{Methods}
\label{sec:methods}

We utilize the von Neumann entanglement entropy \(\mathcal{S}_{\mathrm{vN}}\) to map out the phase diagram, as it provides a sensitive probe of both quantum criticality and symmetry-protected topological structure. For a quantum many-body system in the ground state \(\ket{\Psi_0}\), bipartitioned into subsystems \(\mathcal{A}\) and \(\mathcal{B}\), the bipartite entanglement entropy is defined as
\begin{eqnarray}
\label{eq:vonee}
S_{\mathrm{vN}} = -\operatorname{Tr} \rho_{\mathcal{A}} \ln \rho_{\mathcal{A}},
\end{eqnarray}
where \(\rho_{\mathcal{A}} = \operatorname{Tr}_{\mathcal{B}}\left(\left|\Psi_0\right\rangle\left\langle\Psi_0\right|\right)\) is the reduced density matrix of subsystem \(\mathcal{A}\). Denoting the eigenvalues of \(\rho_{\mathcal{A}}\) by \(\lambda_k\) \((k=1,2,3,\ldots)\) in descending order, we define ES as \(\varepsilon_k=-\ln \lambda_k\). In one-dimensional gapped phases, \(\mathcal{S}_{\mathrm{vN}}\) obeys an area law and saturates to a constant with increasing system size \cite{EisertColloquium2010RMP}. In the SPT$_1$ and SPT$_2$ phases, we observe \(\mathcal{S}_{\mathrm{vN}}\) to be close to \(\ln 2\) and \(2\ln 2\), respectively, consistent with the associated edge multiplets and ES degeneracy structures. At quantum critical points or within extended gapless phases described by conformal field theory (CFT), $S_{\rm vN}$ exhibits universal logarithmic scaling \cite{HolzheyGeometric1994NPB,VidalEntanglement2003PRL,PasqualeCalabreseEntanglement2004JSM,CalabreseEntanglement2009JPAMT}. For one-dimensional systems with OBCs, it follows
\begin{eqnarray}
\label{eq:cfteeform}
S_{\mathrm{vN}}
= \frac{c}{6} \ln \left\{\frac{4(L+1)}{\pi}
\sin \left[\frac{\pi(2l+1)}{2(L+1)}\right]\right\}
	+	s_o,
\end{eqnarray}
where \(c\) is the central charge and \(s_o\) is a nonuniversal constant. In our calculations, \(\mathcal{A}\) contains half of the unit cells, corresponding to an inter-cell cut at the center of the chain. Accordingly, \(S_{\mathrm{vN}}\) grows logarithmically with system size \(L\) at critical points or within gapless phases, while it saturates in gapped phases, enabling us to identify critical lines and gapless regions in the phase diagram. We extract the central charge by fitting the mid-chain entanglement entropy to Eq.~\eqref{eq:cfteeform}.

The three symmetry-breaking phases in our model are characterized by the order parameters
\begin{eqnarray}
\label{eq:cdwm2}
\hat{M}_2 = \frac{1}{L}\sum_i \left(-1\right)^{i} \hat{N}_i
\end{eqnarray}
with \(\hat{N}_i = \hat{n}_{A,i} + \hat{n}_{B,i}\) for the CDW$_2$ phase,
\begin{eqnarray}
\label{eq:cdwm4}
\hat{M}_4 = \frac{1}{L}\sum_i \left(\hat{N}_{4i-3}+\hat{N}_{4i}-\hat{N}_{4i-2}-\hat{N}_{4i-1}\right)
\end{eqnarray}
for the CDW$_4$ phase, corresponding to the pattern in which the first and fourth unit cells within each four-cell block are preferentially occupied, and
\begin{eqnarray}
\label{eq:spm-}
\hat{M}_- = \frac{1}{L} \sum_i \left( \hat{n}_{A,i} - \hat{n}_{B,i} \right)
\end{eqnarray}
for the SP phase. To locate the phase transitions and extract the correlation-length exponent \(\nu\), we compute the Binder cumulant \cite{BinderMonte1985JoCP,ZhangProbing2025NC,LiaoPhase2025PRB}
\begin{eqnarray}
\label{eq:bindercumulant}
U_4 = 1 - \frac{\langle \hat{M}^{4} \rangle}{3 \langle \hat{M}^{2} \rangle^{2}}.
\end{eqnarray}
According to finite-size scaling theory, near a continuous phase transition point the Binder cumulant obeys the scaling form
\begin{equation}
\label{eq:U4universalfunction}
U_4 = f \left( L^{1/\nu} \delta \right),
\end{equation}
where \(\delta\) denotes the distance to the critical point and \(f(x)\) is a universal function. In practice, we evaluate \(U_4\) in the vicinity of the crossing point for different system sizes \(L\) and perform a data collapse by fitting \(U_4\) to a polynomial function of \(x_{L,t}=L^{1/\nu}\left(t - t_c\right)\), where \(t\) is the control parameter driving the transition. The critical point \(t_c\) and the exponent \(\nu\) are determined by minimizing the sum of squared residuals in the data-collapse procedure.

The three gapless phases in our system are described by Luttinger-liquid theory, and the transitions out of these phases are of Berezinskii-Kosterlitz-Thouless (BKT) type. In the vicinity of a BKT transition, but on the gapped side, the correlation length diverges as \(\xi \sim \exp(b/\sqrt{\delta})\), where \(b\) is a nonuniversal constant. Correspondingly, the finite-size scaling of the excitation gap follows \cite{WallinResistance1995PRB,MishraPhase2011PRB,ZhangTruncation2021PRB}
\begin{eqnarray}
\label{eq:dEuniversalfunction}
L \Delta E \left(1 + \frac{1}{2 \ln L + C}\right) = g(x_{L,t}),
\end{eqnarray}
where \(g(x)\) is a universal scaling function, \(C\) is a nonuniversal constant, and \(x_{L,t} = \ln(L/\xi) = \ln L - b/\sqrt{|t-t_c|}\). Here \(\Delta E\) denotes the neutral gap within the half-filling sector, while the charge gaps associated with adding or removing fermions are defined as \(\Delta_1=E(L+1)+E(L-1)-2E(L)\) and \(\Delta_2=E(L+2)+E(L-2)-2E(L)\), where \(E(N)\) is the ground-state energy in the \(N\)-particle sector. We determine the phase boundaries of the gapless phases by performing data collapse based on this scaling ansatz. In addition, the Luttinger parameter \(K\) is extracted from correlation functions. In LL, the single-particle and two-particle correlation functions decay algebraically as
\begin{eqnarray}
\label{eq:singlecorr}
\langle \hat{c}^\dagger_{\alpha,i} \hat{c}_{\beta,j} \rangle \sim |i-j|^{-\frac{1}{2}(K+1/K)}, \\
\label{eq:paircorr}
\langle \hat{c}^\dagger_{A,i} \hat{c}^\dagger_{B,i} \hat{c}_{B,j} \hat{c}_{A,j} \rangle \sim |i-j|^{-2/K}.
\end{eqnarray}
In the Luther-Emery-type gapless phases, the single-particle correlations are short ranged, while the dominant pair correlations behave as \cite{GiamarchiQuantum2003}
\begin{eqnarray}
\label{eq:lecorrs}
\langle \hat{c}^\dagger_{A,i} \hat{c}^\dagger_{B,i} \hat{c}_{B,j} \hat{c}_{A,j} \rangle \sim |i-j|^{-1/2K_{\rm{pair}}}.
\end{eqnarray}

To diagnose the topological structure, we compute nonlocal string order parameters (SOPs) in the unified form
\begin{eqnarray}
\label{eq:stringall}
\mathcal{O}^{X}(i,j)
=\Big\langle \hat{Q}^{X}_{i}
\Big(\prod_{m=i+1}^{j-1}\hat{P}_{A,m}\hat{P}_{B,m}\Big)
\hat{Q}^{X}_{j}\Big\rangle,
\end{eqnarray}
where \(\hat{P}_{A/B,m} \equiv \exp\left(\mathrm{i}\pi \hat{n}_{A/B,m}\right)\) is the local parity operator, \(X=\text{0},\text{I},\text{II},J\), and the endpoint operators \(\hat{Q}^{X}_{i}\) and \(\hat{Q}^{X}_{j}\) are chosen such that the string cuts the corresponding dominant bond of each phase. For the trivial phase we use endpoint operators that cut the intra-cell \(t_0\) bond,
\begin{eqnarray}
\nonumber \hat{Q}^{0}_{i}=\left(\hat{n}_{B,i-1}+\hat{n}_{A,i}-1\right)\hat{P}_{B,i}, \\
\hat{Q}^{0}_{j}=\hat{P}_{A,j}\left(\hat{n}_{B,j}+\hat{n}_{A,j+1}-1\right).
\end{eqnarray}
For the SPT$_1$ phase we use endpoint operators that cut the NN inter-cell \(t_1\) bond,
$$\hat{Q}^{\rm I}_{i}=\hat{n}_{A,i}+\hat{n}_{B,i}-1,
\quad
\hat{Q}^{\rm I}_{j}=\hat{n}_{A,j}+\hat{n}_{B,j}-1.$$
Notice that \(\mathcal{O}^{0}(i,j)\) also acquires a finite value in the SPT$_2$ phase because the endpoint operators also cut the \(t_2\) bonds. To distinguish the trivial phase from SPT$_2$, we remove the density operator that connects the string through the \(t_0\) bond and use
$$\hat{Q}^{\rm II}_{i}=\left(\hat{n}_{B,i-1}-\frac{1}{2}\right)\hat{P}_{B,i}, \quad \hat{Q}^{\rm II}_{j}=\hat{P}_{A,j}\left(\hat{n}_{A,j+1}-\frac{1}{2}\right)$$
which probes the NNN inter-cell \(t_2\) bond. For the gSPT phase, the edge degrees of freedom carry current \(\hat{J}_{A/B,i} = \mathrm{i} ( \hat{c}^\dagger_{A/B,i}\hat{c}_{A/B,i+1}-\mathrm{h.c.})\), and we take
$$\hat{Q}^{\text{J}}_{i}
= \hat{J}_{A,i-1}
 \hat{P}_{B,i-1}\hat{P}_{B,i}, \quad \hat{Q}^{\text{J}}_{j}
= \hat{P}_{A,j}\hat{P}_{A,j+1}
 \hat{J}_{B,j}.$$
We evaluate these correlators at large separations \(|i-j|\) to distinguish the trivial, SPT$_1$, SPT$_2$, and gSPT regimes. These SOPs are constructed according to the underlying bonding structure and become finite inside the corresponding phases. As discussed below, they play the role of Haldane-type SOPs, where long-range string order signals a hidden symmetry-breaking pattern \cite{DenNijsPreroughening1989PRB,PollmannSymmetry2012PRB}. We further use edge modes and ES degeneracy structures to characterize these SPT phases.

\subsection{DMRG Parameters}

We perform finite-size density matrix renormalization group (DMRG) calculations \cite{WhiteDensity1992PRL,WhiteDensitymatrix1993PRB} based on matrix product states (MPS) \cite{OstlundThermodynamic1995PRL} to obtain the ground state and low-lying excited states of the system. The code is implemented using the ITensor Julia library with \(U(1)\) symmetry \cite{FishmanITensor2022SPC}. In the eigenstate searches, we gradually increase the maximal bond dimension \(D\) during the variational sweeps until the truncation error \(\epsilon\) falls below \(10^{-10}\) or the maximal bond dimension \(D_m = 800\) is reached. Calculations of critical properties require larger bond dimensions, which are specified where relevant. The DMRG sweeps are terminated once the change in the ground-state energy is smaller than \(10^{-11}\) and the von Neumann entanglement entropy \(S_{\mathrm{vN}}\) changes by less than \(10^{-8}\) between the final two sweeps. System sizes are chosen to be multiples of \(4\), consistent with the period-4 CDW\(_4\) phase. In practice, several tens of sweeps are sufficient for convergence in gapped phases, whereas critical points typically require hundreds of sweeps.

\section{Results} \label{sec:results}

\begin{figure}[t]
\centering 
\includegraphics[width=1\linewidth]{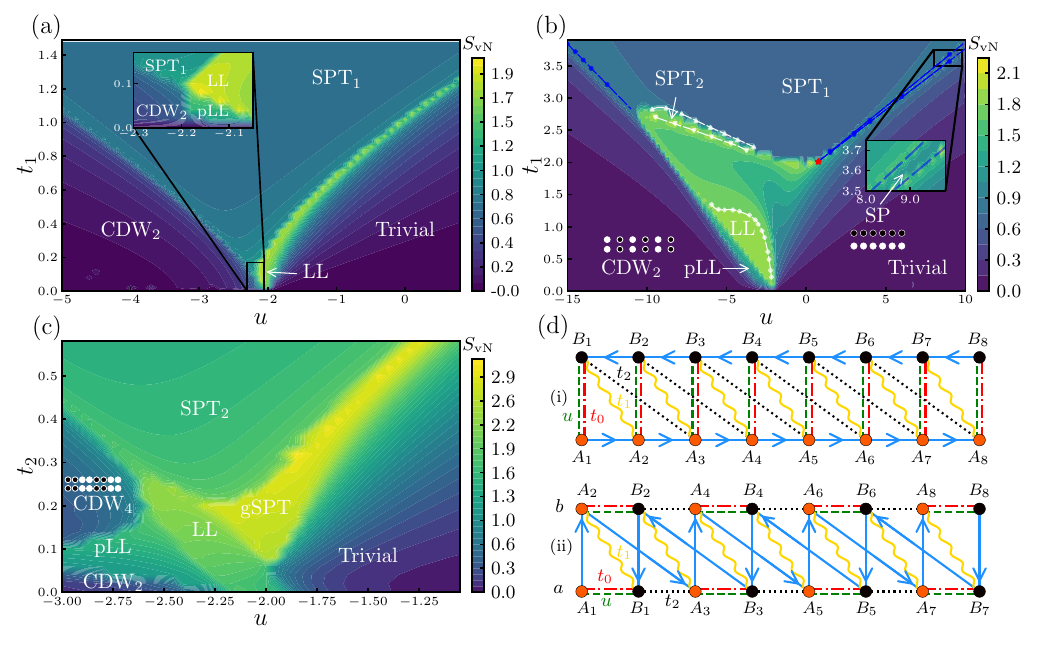}
\caption{Ground-state phase diagrams of the interacting (extended) SSH model. The color scale in (a)–(c) represents \(S_{\rm vN}\) for systems with \(L=256\) under OBC.
(a) For \(t_2=0\), the SPT\(_1\) phase at large \(t_1\) lies between CDW\(_2\) and the trivial phase. At small \(t_1\), a Luttinger-liquid (LL) phase emerges from the vicinity of \(u=-2\), separating the trivial and SPT\(_1\) phases. This LL region shrinks with increasing \(t_1\) and disappears near \(u\approx -1.76\) and \(t_1\approx 0.25\). The inset shows that a paired Luttinger-liquid (pLL) phase appears between the CDW\(_2\) and LL phases near \(u=-2\).
(b) Along \(t_1=2t_2\), SPT\(_1\) is stabilized at large \(t_1\) between the CDW\(_2\) and trivial phases, with a narrow sublattice-polarized (SP) region intervening between SPT\(_1\) and the trivial phase. The SP phase vanishes at about $(t_1,u)=(0.76,2.01)$ (red pentagram). At small \(t_1\), an LL phase again emerges from the vicinity of \(u=-2\), but extends toward more negative \(u\). The pLL phase also separates the CDW\(_2\) and LL phases. In addition, a narrow SPT\(_2\) region appears between SPT\(_1\) and the trivial phase for \(-10 \lesssim u \lesssim -2\).
(c) For fixed \(t_1=0.1\), SPT\(_2\) is stabilized at large \(t_2\) between CDW\(_4\) and the trivial phase. At small \(t_2\), an LL phase emerges near \(u=-2\), separating CDW\(_2\) and the trivial phase. At intermediate \(t_2\), this LL region is flanked by the pLL phase on the more negative-\(u\) side and by the gapless SPT (gSPT) phase on the less negative-\(u\) side.
(d) Schematic current texture of the gSPT ground state. The current-carrying state is constructed from the linear combination \((\psi_1+\mathrm{i}\psi_2)/\sqrt{2}\), where \(\psi_{1,2}\) are the two lowest real-valued ground states obtained from DMRG. The upper panel shows the current texture in the original lattice representation, with current flowing to the right on the \(A\) sublattice and to the left on the \(B\) sublattice. The lower panel shows the rearranged geometry, with odd unit cells placed in the bottom row (new \(a\) sublattice) and even unit cells placed in the top row (new \(b\) sublattice), where the current follows two paths corresponding to the two Luttinger liquids in the original representation.
} 
\label{fig:phasediagram1}
\end{figure}

\subsection{Global phase diagram}
\label{subsec:phasediagram}

\subsubsection{Interacting SSH limit ($t_2 = 0$)}

We first consider the limit \(t_2\to 0\), where the model reduces to the interacting SSH chain with an intra-cell interaction \(u\). The von Neumann entanglement entropy \(S_{\rm vN}\) in the \((u,t_1)\) plane is shown in Fig.~\ref{fig:phasediagram1}(a). Three gapped phases can be identified: two low-entanglement regions at small \(t_1\), and a regime with \(S_{\rm vN}\sim \ln 2\) at larger \(t_1\). For large \(t_1\), inter-cell bonds dominate, producing a topological phase with edge modes and opposite excess charges \(\pm 1/2\) at the two boundaries, corresponding to SPT\(_1\). At small \(t_1\) and large positive \(u\), double occupation within each unit cell is suppressed while the intra-cell bonding favored by \(t_0\) remains intact, resulting in a trivial phase. For large negative \(u\), double occupation is energetically favored. A finite \(t_1\), however, penalizes adjacent doubly occupied cells through virtual hopping, generating an effective repulsion that stabilizes a CDW\(_2\) phase that breaks translational \(\mathbb{Z}_2\) symmetry. These three gapped phases meet near \(t_1\to 0\) and \(u=-2\).

Around this point, a gapless LL phase emerges between the trivial phase and SPT\(_1\). It is identified by two crossing points of the rescaled gap \(L\Delta E\) for different system sizes, with \(L\Delta E\) nearly collapsing onto a common curve between the two crossings, indicating a finite critical regime. From the crossing between the two largest system sizes, \(L=640\) and \(L=768\), we estimate that the LL phase shrinks and disappears near \(u\approx -1.76\) and \(t_1\approx 0.25\). For small \(t_1\lesssim 0.1\), the signatures of SPT\(_1\) disappear and a narrow pLL phase appears between CDW\(_2\) and LL, indicating that SPT\(_1\) terminates at finite \(t_1\) rather than extending all the way to the point \((t_1,u)=(0,-2)\).

The special role of \(u=-2\) can be understood in the limit \(t_1\to 0\) with \(t_0=1\). In this limit, the product state with one bonding fermion per unit cell has energy \(-L\), while a configuration in which half of the unit cells are doubly occupied and the other half empty has energy \(uL/2\). The two therefore become degenerate at \(u=-2\). For \(t_1=0\) and \(u<-2\), the ground state belongs to the latter manifold and is macroscopically degenerate. A finite \(t_1\) lifts this degeneracy through an order-by-disorder mechanism and selects the CDW\(_2\) phase. Near this point, the three lowest states in each unit cell, \(\ket{00}\), \((\ket{10}-\ket{01})/\sqrt{2}\), and \(\ket{11}\), form an effective spin-1 manifold. The \(u\) and \(t_0\) terms generate an onsite anisotropy \((t_0+u/2)\sum_i (S_i^z)^2\) at half filling, so the effective single-ion anisotropy \(D\sim t_0+u/2\) changes sign at \(u=-2\). The inter-cell hopping \(t_1\) generates an XX exchange at leading order, while higher-order processes produce additional anisotropic couplings, resulting in an effective spin-1 chain with predominantly XXZ-type interactions. Under this mapping, the trivial, LL, pLL, SPT\(_1\), and CDW\(_2\) phases correspond to the large-\(D\), gapless XY\(_1\), gapless XY\(_2\), Haldane, and antiferromagnetic phases of the spin-1 chain, respectively, consistent with Ref.~\cite{ChenGroundstate2003PRB}. At \(u=-2\), the single-ion anisotropy vanishes, so the leading-order effective model reduces to a spin-1 XX chain at a BKT point \cite{ChenGroundstate2003PRB,ZhangTruncation2021PRB}. Small deviations away from this point retain strong exchange relative to the anisotropy, which explains the finite LL region around \(u\approx -2\). Moving to slightly more negative \(u\), pair-hopping processes become more important and give the gapless XY\(_2\) regime, corresponding to the pLL phase, which can also be viewed as an effective spin-1/2 XX chain built from the empty and fully occupied unit-cell states. For still more negative \(u\), the negative-\(D\) tendency favors \(|S^z|=1\) on every site, and the longitudinal exchange selects the antiferromagnetic phase, corresponding to CDW\(_2\).

\subsubsection{Extended model with $t_1=2t_2 > 0$}

We now turn on \(t_2>0\) with a fixed ratio \(t_1=2t_2\) and plot \(S_{\rm vN}\) in the \((u,t_1)\) plane in Fig.~\ref{fig:phasediagram1}(b). As in the \(t_2=0\) case, a gapless regime emerges from the special point \((u,t_1)=(-2,0)\), consistent with the effective spin-1 picture in which the single-ion anisotropy vanishes and the leading model is critical. Away from this point, however, the phase diagram differs qualitatively from the \(t_2=0\) case. When \(t_2=0\), increasing \(t_1\) rapidly favors inter-cell bonding and drives the system into SPT\(_1\), while a small negative deviation from \(u=-2\) quickly locks the density into the CDW\(_2\) pattern. As a result, the gapless region remains confined to a relatively narrow range of \(u\) and \(t_1\). Along \(t_1=2t_2\), finite \(t_2\) introduces an additional longer-range hopping channel that enhances kinetic fluctuations and frustrates the period-2 density order, thereby reducing the energy gain of CDW\(_2\). Consequently, stronger attraction is required before the density can lock into CDW\(_2\), shifting the boundary between CDW\(_2\) and the gapless phases to more negative \(u\) and enlarging the gapless region. The intermediate pLL phase between LL and CDW\(_2\) is likewise expanded. Although large negative \(u\) suppresses single-particle motion, correlated two-particle motion can still propagate with an amplitude of order \(t_{\rm pair}\sim (t_1^2+t_2^2)/|u|\), larger than in the \(t_2=0\) case where correlated transport relies only on NN paths that compete more directly with the CDW\(_2\) pattern.

The upper boundary of the gapless region is also pushed upward when \(t_2\) is introduced, again because the additional hopping paths enhance kinetic fluctuations. Upon further increasing \(t_1\) in the regime \(-5 \lesssim u \lesssim -2\), the LL phase first enters an intermediate regime with relatively lower entanglement before transitioning to SPT\(_1\). The entanglement entropy decreases smoothly toward the low-entanglement trivial phase and saturates rapidly with system size, indicating that this regime remains topologically trivial. Thus, the trivial phase that appears at positive \(u\) extends into the negative-\(u\) region down to about \(u\approx -10\), although with an enhanced \(S_{\rm vN}\). In contrast to the \(t_2=0\) case, where LL connects directly to SPT\(_1\), the system now passes through an intermediate trivial regime before entering SPT\(_1\) as \(t_1\) increases.

We also find that both positive and negative \(u\) require larger \(t_1\) to stabilize SPT\(_1\). These trends reflect the combined effects of \(u\) and \(t_2\). For \(u>0\), repulsion suppresses the formation of intracell doublons and thus disfavors inter-cell bonding, so larger \(t_1\) is needed to stabilize SPT\(_1\). Attractive \(u\) suppresses both types of bonds, but penalizes intra-cell bonding more strongly. This explains why, for \(t_2=0\) in Fig.~\ref{fig:phasediagram1}(a), the trivial phase shrinks and smaller \(t_1\) is needed to reach SPT\(_1\) as \(u\) decreases from \(0\) to \(-2\), whereas once CDW\(_2\) emerges at \(u\lesssim -2\), larger \(t_1\) is again required. When \(t_2>0\), the additional hopping paths allow doublon defects on \(t_1\) links to move and disrupt the nonlocal order required for SPT\(_1\). The edge mode can also spread into the bulk through this extra channel, lowering the energy of a trivial state and pushing the onset of SPT\(_1\) to larger \(t_1\). The concomitant growth of inter-cell correlations accounts for the increase in \(S_{\rm vN}\) in this intermediate trivial regime.

For stronger attraction, \(u\ll -t_0\) with \(t_0=1\), the trivial phase can no longer compete because intra-cell bonding is strongly suppressed. In this regime, small \(t_1\) stabilizes CDW\(_2\), while sufficiently large \(t_1\) breaks doublons and forms inter-cell bonds, driving a transition from CDW\(_2\) to SPT\(_1\). At large positive \(u\), the transition between the trivial phase and SPT\(_1\) becomes two-step, with an intermediate SP phase characterized by unequal densities on the two sublattices. Large positive \(u\) favors single occupation in each unit cell, and virtual hopping processes between neighboring \((A,i)\) and \((B,j)\) sites lower the energy of the configurations
\(
\ket{\begin{smallmatrix}
11\\
00
\end{smallmatrix}}\text{ and }
\ket{\begin{smallmatrix}
00\\
11
\end{smallmatrix}}.
\)
This generates an effective ``ferromagnetic'' interaction between neighboring unit cells and stabilizes the SP phase. Such a phase also exists when \(t_2=0\), but only at much larger interaction strength and therefore does not appear in Fig.~\ref{fig:phasediagram1}(a). Near the noninteracting multicritical point at \(u=0\) and \(t_1=2t_2\), the system is known to flow to a \(c=2\) conformal field theory upon introducing either attractive or repulsive interactions \cite{WangQuantum2023PRB}. We also observe evidence for a narrow SPT\(_2\) region between the trivial phase and SPT\(_1\) for \(-10 \lesssim u \lesssim -3\). These regimes involve more complex critical behavior and are not the focus of the present work.

\begin{figure*}[t]
\centering 
\includegraphics[width=1\linewidth]{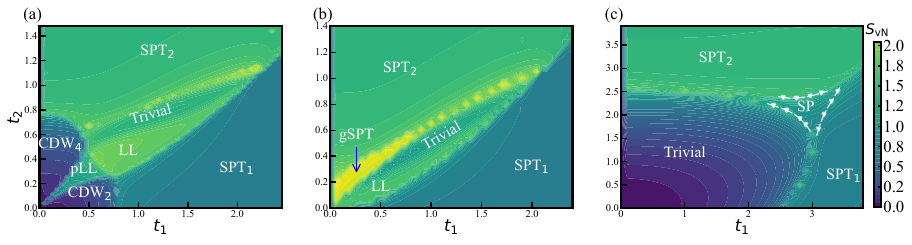}
\caption{Ground-state phase diagrams of the interacting generalized SSH model in the \((t_1,t_2)\) plane for fixed interaction strengths \(u=-4\), \(-2\), and \(6\).
(a) For \(u=-4\), the SPT\(_1\) phase occupies the large-\(t_1\) region, while the SPT\(_2\) phase appears at large \(t_2\). Between them lie a trivial insulating phase adjacent to SPT\(_2\) and an LL phase adjacent to SPT\(_1\). For small \(t_1\) and \(t_2\), strong attractive interactions induce translational-symmetry-breaking charge-density-wave phases: CDW\(_2\) for \(t_1 \gtrsim t_2\) and CDW\(_4\) for \(t_1 \lesssim t_2\). Between these two density-wave phases, a pLL phase emerges and connects continuously to the LL region.
(b) For \(u=-2\), symmetry-breaking phases are absent. Instead, a gSPT phase appears between SPT\(_2\) and the trivial insulating phase, while an LL phase separates the trivial phase from SPT\(_1\).
(c) For \(u=6\), strong repulsive interactions stabilize a symmetry-breaking SP phase, which intervenes among the three topologically distinct insulating phases.} 
\label{fig:phasediagram2}
\end{figure*}

\subsubsection{Phase diagram with $t_1=0.1$ and $t_2 > 0$}
\label{sec:phasediagt10p1}

We have seen that turning on a small \(t_2\) enlarges the gapless regime and also favors the trivial region by destabilizing SPT\(_1\). To investigate the effects of larger \(t_2\), we plot \(S_{\rm vN}\) in the \((u,t_2)\) plane with fixed \(t_1=0.1\) in Fig.~\ref{fig:phasediagram1}(c). This cut intersects the gapless regime in Fig.~\ref{fig:phasediagram1}(a) and allows us to study the evolution of the LL phase by varying only \(t_2\). As discussed above, \(t_2\) enhances kinetic exchange and provides additional paths for doublon motion, so both the LL and pLL regimes are enlarged and extend to more negative \(u\) as \(t_2\) increases from zero. The value of \(S_{\rm vN}\) in the pLL phase is lower than that in the adjacent LL phase because the formation of mobile doublons effectively reduces the number of independent mobile degrees of freedom. For \(t_2\gtrsim 0.1\), a regime with higher entanglement than the LL phase appears on the right side of the LL region around \(u\approx -2\). As will be discussed in the following sections, this regime is a gapless topological phase with doubly degenerate ES, so \(S_{\rm vN}\) is enhanced relative to the nearby LL phase by the additional entanglement associated with the topological sector. As \(t_2\) increases further, the gSPT region first expands and then shrinks, eventually disappearing near \(u\approx -1.7\) and \(t_2\approx 0.3\). For large \(t_2\), NNN unit-cell bonds dominate and the system enters SPT\(_2\), which supports two boundary modes per edge and a richer edge-charge structure than SPT\(_1\). For more negative interaction, \(u\lesssim -2.7\), doublon motion is suppressed and the pLL regime shrinks. In this regime, intermediate \(t_2\) mainly induces CDW\(_2\) order within each odd- and even-unit-cell subchain, and the full system correspondingly forms CDW\(_4\) with a period-4 density pattern.

The appearance of the gSPT phase can be understood from the coupling between two preexisting Luttinger liquids. In the limit of small \(t_1\), the system effectively decomposes into two interacting SSH subchains formed by odd and even unit cells. Reordering the sites gives the two-row geometry shown in Fig.~\ref{fig:phasediagram1}(d). Within each subchain, the hopping \(t_2\) plays the role of the NN inter-cell hopping, so near \(u\approx -2\) each subchain lies in the gapless LL regime, as indicated by Fig.~\ref{fig:phasediagram1}(a). Turning on a small \(t_1\) then hybridizes these two already gapless LLs in a shifted manner, since the \(i\)th site of the odd-cell subchain is coupled to the \((i-1)\)th site of the even-cell subchain. This offset coupling imposes a nontrivial relative phase between the two low-energy sectors. The phase difference is reflected in nonzero imaginary parts of the inter-subchain bilinears, giving the counter-propagating edge-current pattern observed numerically between sites on different subchains, which in the original lattice appears within the \(A\) and \(B\) sublattices [see upper panel in Fig.~\ref{fig:phasediagram1}(d)]. In this way, the glided coupling reorganizes the two LLs into a phase that retains a gapless bulk mode while developing a nontrivial boundary current texture, leading to the gSPT phase. As shown in Fig.~\ref{fig:phasediagram1}(d), the reorganized geometry also suggests a complementary relation between the gapless channels and the current texture: in one representation the LLs reside on the odd- and even-unit-cell sectors and the induced current appears on the horizontal links, while in the other representation the LLs reside on the horizontal links and the induced current appears in the odd- and even-unit-cell sectors.

\subsubsection{Phase diagrams at fixed $u$}

\begin{figure*}[t]
\centering 
\includegraphics[width=1\linewidth]{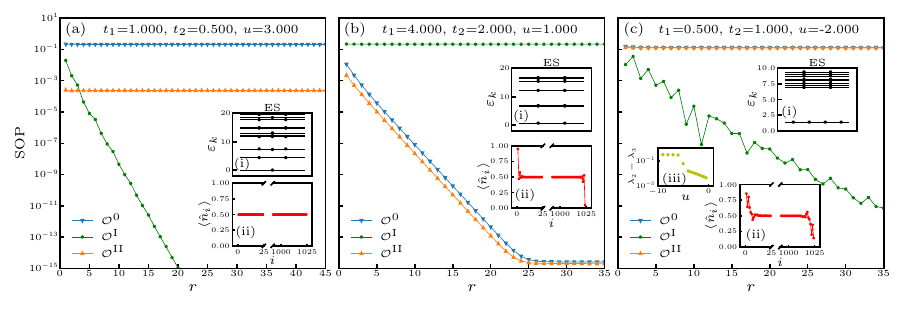}
\caption{String order parameters, edge modes, and entanglement spectra (ES) for three topologically distinct gapped phases.
(a) Trivial phase. Among the string order parameters defined in Eq.~\eqref{eq:stringall}, \(\mathcal{O}^{\rm 0}\) remains finite at long distance, while \(\mathcal{O}^{\rm I}\) decays exponentially to zero and \(\mathcal{O}^{\rm II}\) decays to a value more than two orders of magnitude smaller than \(\mathcal{O}^{\rm 0}\). The inset shows a uniform bulk density profile and a nondegenerate ES.
(b) SPT\(_1\) phase. Only \(\mathcal{O}^{\rm I}\) approaches a finite value, while the other string correlators decay exponentially to zero. The density profile exhibits inversion-related edge modes, and the ES displays an exact two-fold degeneracy.
(c) SPT\(_2\) phase. In contrast to (b), both \(\mathcal{O}^{\rm 0}\) and \(\mathcal{O}^{\rm II}\) remain finite. The ES exhibits an exact two-fold degeneracy, together with a proximate four-fold low-lying structure. Inset (iii) shows that the gap between the lowest two ES levels and the next two levels increases with \(|u|\), but remains small throughout the SPT\(_2\) phase.} 
\label{fig:gappedspt}
\end{figure*}

To further illustrate the effects of interaction, we plot \(S_{\rm vN}\) in the \((t_1,t_2)\) plane at fixed \(u=-4\), \(-2\), and \(6\) in Fig.~\ref{fig:phasediagram2}(a), (b), and (c), respectively. These cuts show how interactions deform the noninteracting phase diagram. For strong attraction, \(u=-4\), doublons on the \(t_0\) links are favored, producing two low-entanglement regimes at small \(t_1\) and \(t_2\): CDW\(_2\) for \(t_1 \gtrsim t_2\) and CDW\(_4\) for \(t_2 \gtrsim t_1\). Between them, doublon motion is enhanced and a pLL phase emerges, consistent with the behavior seen in Fig.~\ref{fig:phasediagram1}(c). Since negative \(u\) suppresses the intra-cell \(t_0\)-bonded structure that stabilizes the trivial phase, the trivial region shrinks for both \(u=-4\) and \(u=-2\). The competing inter-cell-bonded SPT\(_1\) and SPT\(_2\) phases are therefore stabilized over a broader parameter range and expand into part of the original trivial region. Around the original noninteracting critical lines, interactions also generate scattering processes that suppress simple local locking and instead stabilize extended critical regimes \cite{GiamarchiQuantum2003}. As a result, for both \(u=-4\) and \(u=-2\), a finite LL region appears between the trivial phase and SPT\(_1\), replacing the direct transition of the noninteracting case. We do not, however, observe an extended ordinary LL phase between the trivial and SPT\(_2\) phases. A likely reason is that the trivial--SPT\(_2\) boundary lies either at relatively large \(t_1\) and \(t_2\), where stronger attraction may be required to stabilize such a critical regime, or in parameter regions already preempted by CDW\(_4\), as for \(u=-4\), or by gSPT, as for \(u=-2\). In particular, Fig.~\ref{fig:phasediagram2}(b) shows that at \(u=-2\) the gSPT phase appears as soon as a small \(t_1\) is turned on in the small-\(t_2\) regime, consistent with the glided coupling of two preexisting LLs discussed in Sec.~\ref{sec:phasediagt10p1}. At large \(t_1\) and \(t_2\), the trivial phase eventually disappears and SPT\(_1\) and SPT\(_2\) touch directly, as in the noninteracting limit.

For strong repulsion, \(u=6\), doublons are strongly suppressed. This is compatible with the formation of the \(t_0\)-bonded trivial phase, but it disfavors the inter-cell bonding patterns required for SPT\(_1\) and SPT\(_2\), since those states necessarily involve doublon components in superposition. As a result, larger \(t_1\) and \(t_2\) are needed to stabilize the two SPT phases, and the trivial phase is correspondingly enlarged. In the intermediate regime, both NN and NNN inter-cell hoppings induce effective ``ferromagnetic'' interactions between singly occupied unit cells, which stabilize the SP phase visible in Fig.~\ref{fig:phasediagram2}(c) around \(t_1\approx 3\) and \(t_2\approx 2.3\). Reducing either \(t_1\) or \(t_2\) weakens this tendency and causes the SP phase to shrink. On the other hand, when both hoppings become sufficiently large, inter-cell bonding becomes dominant, doublon fluctuations are restored, and the density imbalance of the SP phase is reduced. The SP region therefore shrinks again, and at still larger \(t_1\) and \(t_2\) the two SPT phases meet directly, similar to the noninteracting case.

\subsection{Gapped phases}
\label{subsec:gappedphase}

\begin{figure}[t]
\centering 
\includegraphics[width=1\linewidth]{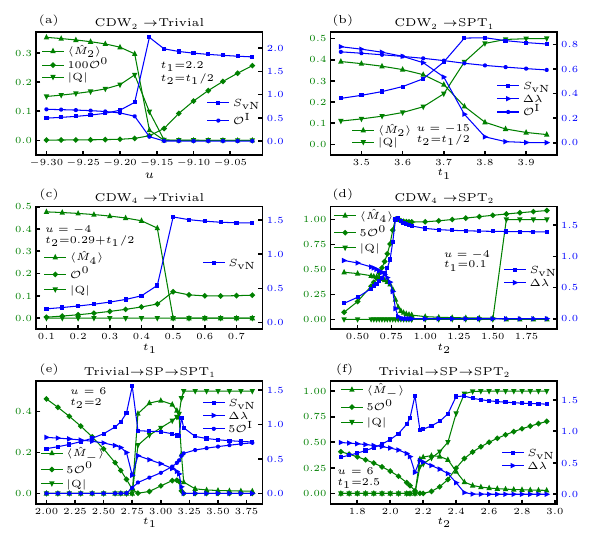}
\caption{Order parameters, \(S_{\rm vN}\), Schmidt gap \(\Delta\lambda\), SOPs, and excess charge \(Q\) across representative transitions between symmetry-breaking and symmetric phases:
(a) CDW\(_2\) to trivial,
(b) CDW\(_2\) to SPT\(_1\),
(c) CDW\(_4\) to trivial,
(d) CDW\(_4\) to SPT\(_2\),
(e) trivial to SP to SPT\(_1\), and
(f) trivial to SP to SPT\(_2\).
The system size is \(L=256\). The corresponding symmetry-breaking order parameters remain finite in the ordered phases and vanish upon entering the symmetric phases. The points where the order parameters drop are consistent with the peaks of \(S_{\rm vN}\). In the ordered phases, \(\Delta\lambda\) remains finite and closes after crossing into the SPT phases. The SOPs evolve smoothly across all phase boundaries and are generally also nonzero in the nearby symmetry-breaking phases. In the SPT phases, the excess charge decreases from its quantized value [\(1/2\) for SPT\(_1\) and \(1\) for SPT\(_2\)] upon entering the symmetry-breaking phases. In the part of the SPT\(_2\) regime close to CDW\(_4\), however, the lowest-energy state lies in an edge sector with zero excess charge, even though \(\Delta\lambda\) remains closed and the SOP stays finite. For clarity, the plotted SOPs in (a), (d), (e), and (f) have been multiplied by numerical factors.
} 
\label{fig:cdwstospts}
\end{figure}

\subsubsection{Gapped symmetry-protected topological phases}

We now discuss the gapped SPT phases. Under OBCs, the distinct gapped phases are diagnosed by complementary bulk and boundary signatures, including nonlocal SOPs, ES, and boundary excess charges. Figure~\ref{fig:gappedspt} summarizes these quantities for the trivial phase, SPT$_1$, and SPT$_2$. Our SPT$_1$ is adiabatically connected, at leading order, to the Haldane phase of the spin-1 XXZ chain with single-ion anisotropy, so it is natural to construct its SOP from the Haldane string operator \cite{PollmannSymmetry2012PRB}. Under the spin-1 mapping, the endpoint operator is \(\hat{S}^z_i=\hat{n}_{A,i}+\hat{n}_{B,i}-1\), while the string part reduces to the parity operator on the corresponding bond, leading to \(\mathcal{O}^{\rm I}\) in Eq.~\eqref{eq:stringall}. Shifting the bond center by one lattice spacing, \((B,i)\rightarrow(A,i)\) and \((A,i)\rightarrow(B,i-1)\), changes the endpoint operator to \(\hat{n}_{B,i-1}+\hat{n}_{A,i}-1\), from which we define \(\mathcal{O}^{\rm 0}\) for the trivial phase. Physically, a finite SOP requires the endpoint operators to cut the underlying bonding structure an odd number of times \cite{AnfusoString2007PRB}; with the appropriate endpoint choice, the endpoint contribution and the string parity combine constructively and yield a nonzero long-distance value. As shown in Figs.~\ref{fig:gappedspt}(a) and (b), \(\mathcal{O}^{\rm 0}\) saturates in the trivial phase and decays exponentially in SPT$_1$, while \(\mathcal{O}^{\rm I}\) saturates in SPT$_1$ and decays exponentially in the trivial phase, so these two SOPs sharply distinguish the two phases.

The situation is different in SPT$_2$. There, the endpoint operator of \(\mathcal{O}^{\rm I}\) cuts the relevant bonding structure twice, so the corresponding contributions cancel and \(\mathcal{O}^{\rm I}\) vanishes. By contrast, \(\mathcal{O}^{\rm 0}\) still cuts the structure once and therefore remains finite in SPT$_2\), as confirmed in Fig.~\ref{fig:gappedspt}(c). To further characterize this phase, we remove the explicit intra-cell \(t_0\) contribution from the endpoint operator and define \(\mathcal{O}^{\rm II}\) using \(\hat{n}_{B,i-1}-1/2\). In SPT$_2$, both \(\mathcal{O}^{\rm 0}\) and \(\mathcal{O}^{\rm II}\) remain finite at long distance. In the trivial phase, however, \(\mathcal{O}^{\rm II}\) is suppressed by several orders of magnitude and retains only a small residual value, because \(\hat{n}_{B,i-1}\) still contains some information from the nearby \(t_0\) bond through the \(t_1\) coupling. Therefore, \(\mathcal{O}^{\rm II}\) is useful for identifying SPT$_2\), although it is not by itself as sharp a discriminator between the trivial phase and SPT$_2$ as \(\mathcal{O}^{\rm 0}\) and \(\mathcal{O}^{\rm I}\) are for the trivial phase and SPT$_1$. Overall, the SOP pattern shows that SPT$_2$ carries a distinct nonlocal correlated structure rather than being a simple continuation of either the trivial phase or SPT$_1$.

A cleaner distinction among the three gapped phases is provided by the ES and the boundary excess charge. As shown in the insets of Fig.~\ref{fig:gappedspt}, the trivial phase has a nondegenerate ES, SPT$_1$ exhibits an exact two-fold degeneracy, and SPT$_2$ also shows an exact two-fold ES degeneracy with a proximate four-fold low-lying structure consisting of two nearby doublets. This can be understood from the fixed-point bonding patterns across the entanglement cut. In the large-\(t_0\) limit, the ground state is a product of intra-cell dimers, so the ES is nondegenerate. In the large-\(t_1\) and large-\(t_2\) limits, the cut breaks one and two inter-cell bonds, respectively, leading to the exact two-fold and four-fold ES degeneracies at the corresponding fixed points. The boundary physics is consistent with this picture. For dominant \(t_1\), one boundary mode appears at each end, giving exponentially localized edge states with excess charge \(Q = \sum_{i=1}^{L/2}\langle \hat{n}_{A,i}+\hat{n}_{B,i}-1\rangle=\pm 1/2\). For dominant \(t_2\), two boundary modes appear at each end, generating a six-fold degenerate edge manifold in the noninteracting limit.

Once interactions are introduced, chiral symmetry is broken while inversion symmetry is preserved. As a result, the exact four-fold ES degeneracy of the noninteracting SPT$_2$ fixed point is reduced to two nearby doublets. The splitting between the two doublets grows with \(|u|\) but remains small throughout the SPT$_2$ phase, as shown in inset (iii) of Fig.~\ref{fig:gappedspt}(c). Interactions also reorganize the edge manifold through symmetry-allowed local couplings between the two boundary modes. The lowest-energy state of SPT$_2$ often carries quantized boundary excess charge \(Q=\pm1\). However, this is not universal throughout the whole phase: in the part of the SPT$_2$ regime close to CDW$_4$, the lowest-energy state can instead have \(Q=0\), as shown in Fig.~\ref{fig:cdwstospts}. Therefore, the excess charge of the lowest state is not by itself a universal diagnostic of SPT$_2$, whereas the exact two-fold ES degeneracy with a proximate four-fold ES structure and the nonzero SOP remain robust characteristics of this phase. In this sense, interacting SPT$_2$ is not simply two decoupled copies of SPT$_1$: symmetry-allowed couplings hybridize the two topological channels, reorganize the edge manifold, and lift the exact free-fermion edge degeneracy, while preserving a distinct interacting topological phase continuously connected to the winding-number-two limit.

\subsubsection{Symmetry-breaking phases}

When the interaction is sufficiently strong, the system develops spontaneous symmetry breaking. As discussed above, the three symmetry-breaking phases are CDW$_2$, CDW$_4$, and SP, characterized by the local order parameters \(M_2\), \(M_4\), and \(M_-\) defined in Eqs.~\eqref{eq:cdwm2}--\eqref{eq:spm-}. CDW$_2$ and CDW$_4$ are favored primarily when \(t_1\) and \(t_2\) dominate, respectively. As a result, CDW$_2$ (CDW$_4$) lies mainly adjacent to SPT$_1$ (SPT$_2$), while direct transitions to SPT$_2$ (SPT$_1$) are strongly suppressed. By contrast, both \(t_1\) and \(t_2\) contribute to the formation of the SP phase, allowing it to interpolate continuously between the two SPT phases.

Figure~\ref{fig:cdwstospts} shows these order parameters along representative cuts across symmetry-breaking and nearby symmetric gapped phases: CDW$_2$ to trivial and SPT$_1$ in Figs.~\ref{fig:cdwstospts}(a) and (b), CDW$_4$ to trivial and SPT$_2$ in Figs.~\ref{fig:cdwstospts}(c) and (d), and trivial to SP to SPT$_1$ and trivial to SP to SPT$_2$ in Figs.~\ref{fig:cdwstospts}(e) and (f), respectively. For comparison with the SPT phases, we also plot \(S_{\rm vN}\), the Schmidt gap \(\Delta\lambda=\lambda_1-\lambda_2\), the SOPs, and the excess charge \(Q\). In each case, the corresponding local order parameter is finite only inside the symmetry-breaking phase and vanishes upon entering the adjacent symmetric phase, confirming that \(M_2\), \(M_4\), and \(M_-\) correctly diagnose the breaking of \(\mathbb{Z}_2\) translation, \(\mathbb{Z}_4\) translation, and \(\mathbb{Z}_2\) inversion symmetry, respectively. The location where the order parameter changes most rapidly nearly coincides with the peak of \(S_{\rm vN}\), consistent with the presence of phase transitions. In all symmetry-breaking phases, \(\Delta\lambda\) remains finite and decreases to zero upon entering the SPT phases. For transitions from symmetry-breaking phases to the trivial phase, the order parameters and \(S_{\rm vN}\) often show an apparently abrupt change. This is a finite-size and finite-bond-dimension effect in DMRG, since the onset of symmetry breaking appears artificially sharp once the correlation length reaches the largest numerically accessible scale. By contrast, for transitions from symmetry-breaking phases to SPT phases, the order parameters, \(S_{\rm vN}\), and \(\Delta\lambda\) evolve smoothly across the transition.

We also examine the long-distance SOPs \(\mathcal{O}^{0}\) and \(\mathcal{O}^{I}\) at separation \(|j-i|=120\) with \(i=L/4+1\). For all cuts, \(\mathcal{O}^{0}\) is nonzero in the trivial and SPT$_2$ phases and vanishes in SPT$_1$, while \(\mathcal{O}^{I}\) is nonzero in SPT$_1$ and vanishes in the trivial and SPT$_2$ phases, consistent with Fig.~\ref{fig:gappedspt}. At the same time, the SOPs evolve smoothly across the boundaries between symmetry-breaking and symmetric phases. In particular, the nonzero \(\mathcal{O}^{0}\) in the trivial and SPT$_2$ phases decreases upon entering the symmetry-breaking phases, where the states are closer to product states and the SOP mainly probes local order. Since the endpoint operator takes values close to zero in the ordered phases, \(\mathcal{O}^{0}\) is correspondingly suppressed. By contrast, for \(\mathcal{O}^{I}\) the endpoint operator also probes the local order but takes a nonzero value, so \(\mathcal{O}^{I}\) remains nonzero in the symmetry-breaking phases and can even exceed its value in the neighboring SPT phase, as seen for CDW$_2$ in Fig.~\ref{fig:cdwstospts}(b). Therefore, while the SOPs are useful for distinguishing topologically distinct symmetric phases, they do not sharply distinguish symmetry-breaking phases from symmetric phases.

Finally, we consider the excess charge \(Q\). In SPT$_1$, \(|Q|\) is quantized to \(1/2\), while the nearby CDW$_2$ and SP phases can still host nonzero but nonquantized values of \(|Q|\) because they do not possess symmetry-protected boundary modes. Correspondingly, \(|Q|\) decreases smoothly when going from SPT$_1$ into CDW$_2$ in Fig.~\ref{fig:cdwstospts}(b) and into SP in Fig.~\ref{fig:cdwstospts}(e). In Fig.~\ref{fig:cdwstospts}(a), the CDW$_2$ phase is also close to SPT$_1$ and therefore still shows nonzero \(|Q|\), but \(|Q|\) quickly drops to zero upon entering the trivial phase, where the dominant \(t_0\) bonding restores inversion symmetry and a uniform density distribution. The behavior of \(Q\) in SPT$_2$ is more subtle. In the noninteracting limit, SPT$_2$ has two boundary modes on each edge and a six-fold degenerate boundary manifold, among which two states have \(Q=\pm1\) while the others have \(Q=0\). Once interactions are turned on, this degeneracy is lifted, and the excess charge of the lowest-energy state depends on the parameter regime. Close to CDW$_4$, the lowest-energy state of SPT$_2$ has \(Q=0\), whereas close to SP it has \(|Q|=1\). This is consistent with the character of the nearby symmetry-breaking phases. In CDW$_4$, a finite \(t_1\) penalizes neighboring occupied unit cells, and for an open chain with \(L\) a multiple of \(4\) the pattern built from repeated blocks \(\ket{\begin{smallmatrix}1001\\1001\end{smallmatrix}}\) contains one fewer NN occupied pair than the competing pattern built from \(\ket{\begin{smallmatrix}1100\\1100\end{smallmatrix}}\), so it is energetically favored, preserves inversion symmetry, and therefore gives zero excess charge, as shown in Fig.~\ref{fig:cdwstospts}(c). By contrast, the SP phase breaks inversion symmetry and favors one sublattice over the other, which is more compatible with the \(Q=\pm1\) boundary sector of SPT$_2$. For sufficiently large \(t_2\), the lowest-energy state in SPT$_2$ recovers \(Q=\pm1\), as shown in Fig.~\ref{fig:cdwstospts}(d), where \(|Q|\) changes from \(0\) to \(1\) near \(t_2\approx 1.5\).

\subsection{Gapless phases}
\label{subsec:gaplessphase}
\begin{figure}[t]
\centering 
\includegraphics[width=1\linewidth]{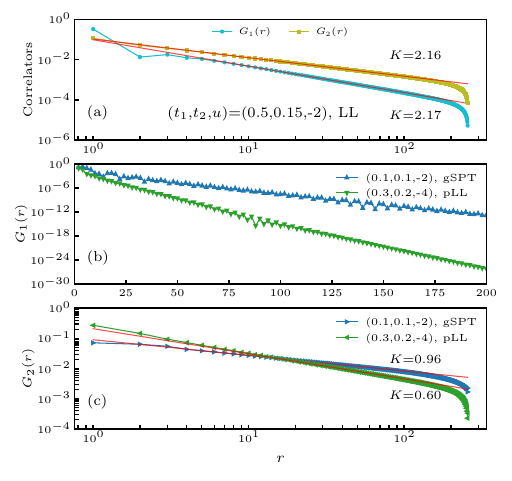}
\caption{Single-particle and two-particle correlation functions \(G_1(r)\) and \(G_2(r)\) in the LL, pLL, and gSPT phases.
(a) In the LL phase, both the single-particle and two-particle correlation functions exhibit algebraic decay. The Luttinger parameter extracted from the two correlators agrees well.
(b) and (c) In both the pLL and gSPT phases, the single-particle correlator decays exponentially, while the two-particle correlator remains algebraic. The Luttinger parameters are extracted from the two-particle correlation functions.}
\label{fig:gaplessphasecorrelators}
\end{figure}

\begin{figure}[t]
\centering 
\includegraphics[width=1\linewidth]{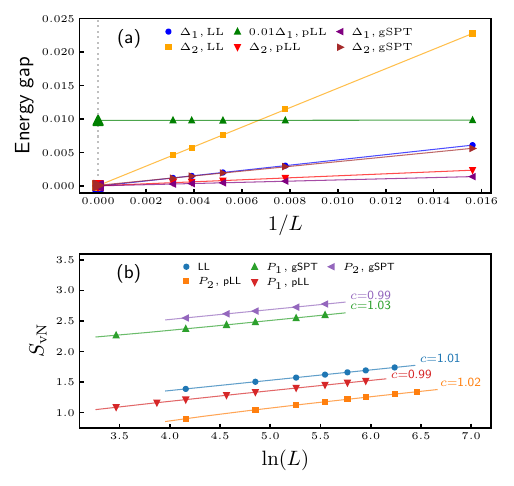}
\caption{Charge gaps and \(S_{\rm vN}\) in the LL, pLL, and gSPT phases.
(a) The charge-1 gap \(\Delta_1\) and charge-2 gap \(\Delta_2\) both extrapolate to zero with increasing system size in the LL phase at \((t_1,t_2,u)=(0.5,0.15,-2)\) and in the gSPT phase at \((0.1,0.1,-2)\). In contrast, in the pLL phase at \((0.3,0.2,-4)\), \(\Delta_2\) extrapolates to zero while \(\Delta_1\) remains finite in the thermodynamic limit.
(b) The value of \(S_{\rm vN}\) scales linearly with \(\ln L\) in all three phases. The data points shown here are \((t_1,t_2,u)=(0.5,0.15,-2)\) for the LL phase, \(P_1=(0.1,0.1,-2.7)\) and \(P_2=(0.3,0.2,-4)\) for the pLL phase, and \(P_1=(0.1,0.1,-2)\) and \(P_2=(0.1,0.18,-2)\) for the gSPT phase. The extracted central charges are consistent with the CFT prediction \(c=1\).}
\label{fig:gaplessphasegaps_svns}
\end{figure}

\begin{figure}[t]
\centering 
\includegraphics[width=1\linewidth]{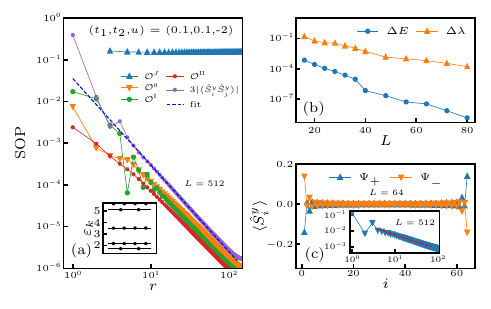}
\caption{SOPs, neutral energy gap \(\Delta E\), Schmidt gap \(\Delta\lambda\), and edge current modes \(\langle \hat{S}^y_i \rangle\) in the gSPT phase at \(t_1=0.1\), \(t_2=0.1\), and \(u=-2\).
(a) The SOP defined with the current endpoint operator saturates to a finite value at long distance, whereas the other SOPs and the current-current correlation function decay algebraically to zero. The results are shown for \(L=512\). A linear fit of \(\ln |\langle \hat{S}^y_i \hat{S}^y_j \rangle|\) versus \(\ln |i-j|\) gives a slope of \(-2.06\). The inset shows the two-fold degeneracy of the ES.
(b) The many-body neutral gap within the half-filled sector and the Schmidt gap both decrease exponentially with system size.
(c) The current-carrying edge modes are obtained from linear combinations of the two degenerate ground states, \(|\Psi_{\pm}\rangle=(|\psi_{1}\rangle \pm \mathrm{i}|\psi_{2}\rangle)/\sqrt{2}\). The inset shows that the magnitude of the edge-current profile decays algebraically into the bulk, with a linear fit giving a slope of \(-0.89\).}
\label{fig:gsptproperties}
\end{figure}

Unless otherwise specified, all calculations of critical properties in this subsection and Sec.~\ref{subsec:phasetransition} use a DMRG bond dimension \(D=1000\).

\subsubsection{Critical properties of the gapless phases}

We now discuss the critical properties of the gapless phases. In addition to the gapped phases discussed above, the phase diagram contains extended regions of LL, pLL, and gSPT behavior. In the LL phase, both the single-particle and pair correlation functions decay algebraically, whereas in the pLL and gSPT phases the single-particle correlator decays exponentially while the pair correlator remains algebraic. Figure~\ref{fig:gaplessphasecorrelators}(a) shows the correlators \(G_1(r)=\langle \hat{c}^\dagger_{B,L/2} \hat{c}_{A,L/2+r}\rangle\) and \(G_2(r)=\langle \hat{c}^\dagger_{A,L/2}\hat{c}^\dagger_{B,L/2} \hat{c}_{B,L/2+r}\hat{c}_{A,L/2+r}\rangle\) in the LL phase at \(t_1=0.5\), \(t_2=0.15\), and \(u=-2\). Both correlators exhibit algebraic decay, and fitting them with Eqs.~\eqref{eq:singlecorr} and \eqref{eq:paircorr} yields consistent Luttinger parameters \(K=2.17\) and \(2.16\), respectively. Figure~\ref{fig:gaplessphasecorrelators}(c) shows that the pair correlator also decays algebraically in the pLL phase at \(t_1=0.3\), \(t_2=0.2\), and \(u=-4\), as well as in the gSPT phase at \(t_1=t_2=0.1\) and \(u=-2\); fitting them to Eq.~\eqref{eq:lecorrs} gives \(K_{\rm pair}=0.60\) and \(0.96\), respectively. By contrast, Fig.~\ref{fig:gaplessphasecorrelators}(b) shows that the single-particle correlator decays exponentially in both the pLL and gSPT phases. For pLL, the physical picture is simple: strong attractive interactions bind the two fermions within a unit cell into doublon-like composite objects, so low-energy transport is dominated by pair fluctuations, while single-particle excitations remain costly. For gSPT, the same exponential decay reflects the pinning of the relative phase between the two particles, which is accompanied by the emergence of current-carrying edge modes, as discussed below.

This distinction is further confirmed by the charge gaps shown in Fig.~\ref{fig:gaplessphasegaps_svns}(a). In the pLL phase, the single-particle gap \(\Delta_1\) extrapolates to a finite value in the thermodynamic limit, whereas the two-particle gap \(\Delta_2\) scales to zero as \(1/L\). In contrast, both \(\Delta_1\) and \(\Delta_2\) vanish as \(1/L\) in the LL and gSPT phases. Thus, although the single-particle correlator decays exponentially in both pLL and gSPT, only pLL has a finite single-particle gap. In gSPT, adding or removing a particle remains gapless because the extra charge can be accommodated by the boundary modes, whose current profile decays algebraically into the bulk. Finally, Fig.~\ref{fig:gaplessphasegaps_svns}(b) shows \(S_{\rm vN}\) as a function of \(\ln L\). For LL at \((t_1, t_2, u) = (0.5, 0.15, -2)\) and pLL at \((0.1,0.1,-2.7)\) and \((0.3,0.2,-4)\), using bond dimension \(D=1000\), the extracted central charges are \(1.01\), \(0.99\), and \(1.02\), respectively, in excellent agreement with the CFT prediction \(c=1\). For gSPT at \((0.1,0.1,-2)\) and \((0.1,0.18,-2)\), the entanglement is substantially larger, and a reliable extraction requires extrapolation to the infinite-bond-dimension limit. We therefore obtain \(S_{\rm vN}\) for each \(L\) by extrapolating data computed with \(D=600,700,\ldots,1200\), using a power-law fit in \(1/D\). The extrapolated central charges, \(c=1.03\) and \(0.99\), are again consistent with the CFT prediction \(c=1\), showing that gSPT also contains only a single gapless mode. In all fittings of \(S_{\rm vN}\), we include a \(1/L\) correction. These results show that all three gapless phases are described by a \(c=1\) CFT, while differing in the nature of their gapless excitations and correlation functions.

\subsubsection{Gapless symmetry-protected topological phase}

We next turn to the topological properties of the gSPT phase. Figure~\ref{fig:gsptproperties}(a) shows all SOPs defined in Eq.~\eqref{eq:stringall} at \(t_1=t_2=0.1\), \(u=-2\), and \(L=512\), with the left endpoint operator fixed at \(i=L/2+1\). We find that \(\mathcal{O}^{\rm 0}\), \(\mathcal{O}^{\rm I}\), and \(\mathcal{O}^{\rm II}\) all decay algebraically to zero, whereas the SOP defined with the current endpoint operator, \(\mathcal{O}^{\rm J}\), saturates to a finite value of about \(0.345\). This shows that the gSPT phase has neither local nor nonlocal density order. Indeed, the density profile is flat in this phase. Instead, the nontrivial order resides in the current operator \(\hat{J}_{A/B}\), or equivalently in the \(y\) component of the effective spin defined in the rearranged geometry of Fig.~\ref{fig:phasediagram1}(d),
\(
\hat{S}^y_i=1/2\sum_{s,s'} \hat{c}^{\dagger}_{s,i}\sigma^y_{ss'}\hat{c}_{s',i},
\)
where \(s,s'=a,b\) label the two rows. We also find that the correlation function \(|\langle \hat{S}^y_i \hat{S}^y_j\rangle|\) decays algebraically to zero, indicating the absence of true long-range local order in \(\hat{S}^y_i\). Therefore, the finite \(\mathcal{O}^{\rm J}\) characterizes a hidden current order and signals the topological nature of the gSPT phase. Further evidence for the topological sector is provided in Fig.~\ref{fig:gsptproperties}(b), where both the neutral gap and the Schmidt gap decrease exponentially with system size.

The finite SOP \(\mathcal{O}^{\rm J}\) also implies that \(\hat{S}^y\) is the charged endpoint operator, so each boundary should carry a spontaneous expectation value \(\langle \hat{S}^y_i\rangle \neq 0\) \cite{ThorngrenIntrinsically2021PRB}. However, a nonzero \(\langle \hat{S}^y_i\rangle\) requires a complex wavefunction, whereas both the Hamiltonian and its MPS/MPO representation in DMRG are real. As a result, DMRG returns real superpositions of the two degenerate ground states, for which \(\langle \hat{S}^y_i\rangle=0\), which is consistent with the two-fold degeneracy of the ES shown in the inset of Fig.~\ref{fig:gsptproperties}(a). To expose the boundary mode, we therefore compute two nearly degenerate ground states \(\psi_1\) and \(\psi_2\) and form the linear combinations \(\Psi_{\pm}=\left(\psi_1 \pm i\psi_2\right)/\sqrt{2}\). The resulting profiles of \(\langle \hat{S}^y_i\rangle\) are shown in Fig.~\ref{fig:gsptproperties}(c). They are inversion symmetric, with opposite nonzero values localized near the two boundaries, and \(\Psi_+\) and \(\Psi_-\) carry opposite signs of \(\langle \hat{S}^y_i\rangle\), consistent with their degeneracy. The inset of Fig.~\ref{fig:gsptproperties}(c) shows that \(|\langle \hat{S}^y_i\rangle|\) decays algebraically into the bulk. A fit of \(\ln |\langle \hat{S}^y_i\rangle|\) versus \(\ln i\) gives an exponent \(0.76\), while fitting \(\ln |\langle \hat{S}^y_i \hat{S}^y_j\rangle|\) versus \(\ln|i-j|\) gives \(2.07\). Both values are consistent with the Luttinger-liquid expectation of exponents \(1\) and \(2\), respectively.

Since \(\hat{S}^y\) measures the relative phase between the two sites in the effective rung, the nonzero endpoint order indicates that the relative sector is pinned. As a consequence, the single-particle correlator \(\langle \hat{c}^{\dagger}_{A/B,i}\hat{c}_{A/B,j}\rangle\) decays exponentially, whereas the pair correlator \(\langle \hat{c}^{\dagger}_{A,i}\hat{c}^{\dagger}_{B,i}\hat{c}_{B,j}\hat{c}_{A,j}\rangle\) remains algebraic because the total charge sector stays gapless. Together with the vanishing charge gaps and central charge \(c\approx 1\) discussed above, these observations confirm that the bulk of the gSPT phase is described by a Luttinger liquid, while its boundaries carry topological current modes. In a gapped SPT, the entanglement cut creates virtual boundaries carrying symmetry-protected edge degrees of freedom, and the ES degeneracy directly reflects this fractionalized edge structure. In the gSPT phase, by contrast, the bulk remains gapless and the characteristic boundary modes are associated with a pinned relative phase and edge currents rather than with a short-range-entangled bond structure. Accordingly, the two-fold degeneracy of the low-lying ES should be understood as reflecting the near-degenerate edge-sector structure and its cat-state superposition, rather than the usual virtual-edge degeneracy of a gapped SPT.

The properties of the gSPT phase are consistent with those of the intrinsically gapless topological phase in the doped Ising-Hubbard chain \cite{ThorngrenIntrinsically2021PRB}, the main difference being that the charged endpoint operator is \(\hat{S}^y\) here rather than \(\hat{S}^z\). In Ref.~\cite{ThorngrenIntrinsically2021PRB}, the gapless topology originates from an anomalous low-energy realization of a \(\pi\) rotation \(R_x\) about the \(x\) axis, which satisfies \(R_x^2=P\) with \(P\) the fermion-parity operator. As a result, the microscopic on-site symmetry is \(\mathbb{Z}_4\), while in the low-energy theory the parity subgroup acts only on the gapped fermions and thus becomes invisible, leaving an effective \(\mathbb{Z}_2\) symmetry realized anomalously. In our model, after defining
\(
\hat{S}^x_i=1/2\sum_{s,s'} \hat{c}^{\dagger}_{s,i}\sigma^x_{ss'}\hat{c}_{s',i},
\)
the rotation \(R_x\) is an exact symmetry only at \(t_1=0\). For nonzero \(t_1\), the system remains invariant under the combined operation \(R_x' = R_x T_a = T_a^{-1}R_x\), where \(T_a\) translates row \(a\) toward the left by two lattice spacings. This combined symmetry still satisfies \((R_x')^2=P\), so the same \(\mathbb{Z}_4\) structure persists. Since the fermions remain gapped degrees of freedom while the parity subgroup acts only on them, the low-energy theory again realizes an effective \(\mathbb{Z}_2\) symmetry anomalously. This explains why the phase is gapless and topological at the same time. In particular, the gSPT phase appears only after two gapless modes are formed, showing that the absence of a full gap is essential rather than incidental.

\subsection{Quantum phase transitions}
\label{subsec:phasetransition}
\begin{figure}[t]
\centering 
\includegraphics[width=1\linewidth]{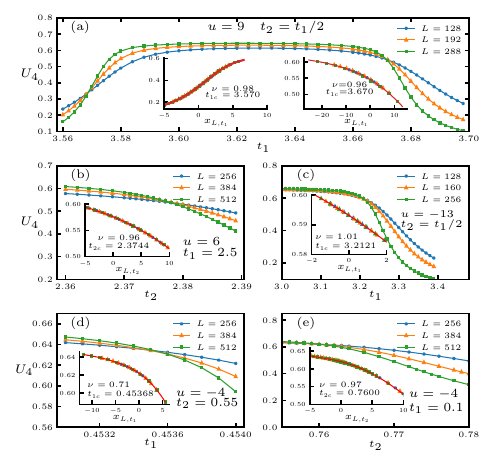}
\caption{Binder cumulant \(U_4\) across representative quantum phase transitions from symmetry-breaking phases to symmetric phases:
(a) trivial to SP to SPT\(_1\),
(b) SP to SPT\(_2\),
(c) CDW\(_2\) to SPT\(_1\),
(d) CDW\(_4\) to trivial, and
(e) CDW\(_4\) to SPT\(_2\).
The insets show the optimal data collapse obtained by tuning the critical point and the correlation-length exponent \(\nu\).}
\label{fig:cdwtransitions}
\end{figure}

\begin{figure}[t]
\centering 
\includegraphics[width=1\linewidth]{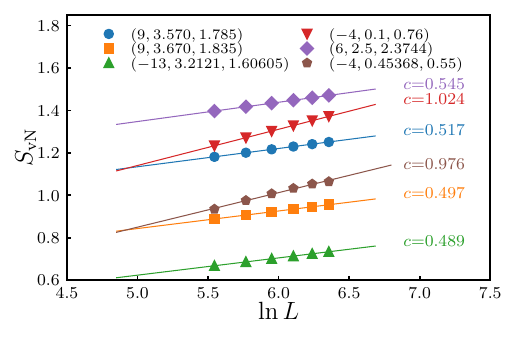}
\caption{Entanglement entropy as a function of \(\ln L\) at the critical points of the symmetry-breaking transitions determined from the Binder-cumulant analysis in Fig.~\ref{fig:cdwtransitions}. The extracted central charges are consistent with \(c=0.5\) for transitions out of CDW\(_2\) and SP, and with \(c=1\) for transitions out of CDW\(_4\).}
\label{fig:cdwtransitionsvns}
\end{figure}

\begin{figure}[t]
\centering 
\includegraphics[width=1\linewidth]{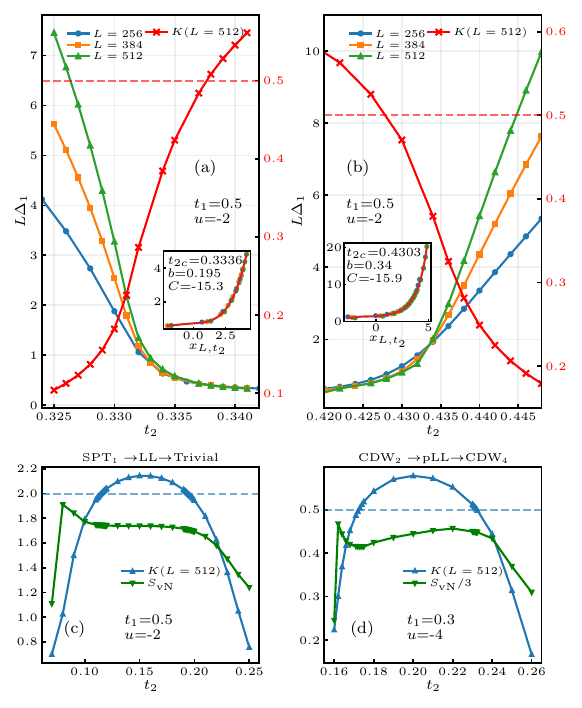}
\caption{Energy gaps and Luttinger liquid parameters across BKT transitions out of gapless phases.
(a) Rescaled charge-1 gap \(L\Delta_1\) as a function of \(t_2\) at \(u=-2\) and \(t_1=0.5\), illustrating the BKT transitions from trivial to gSPT to SPT\(_2\). The insets show the optimal data collapse. The Luttinger parameter \(K\), extracted from correlation functions for \(L=512\), is plotted on the right \(y\) axis.
(b) Luttinger parameter \(K\) in the smaller-\(t_2\) regime along \(u=-2\) and \(t_1=0.5\), corresponding to the BKT transitions from SPT\(_1\) to LL to trivial.
(c) Luttinger parameter \(K\) at \(u=-4\) and \(t_1=0.3\), corresponding to the BKT transitions from CDW\(_2\) to pLL to CDW\(_4\).}
\label{fig:bkttransitions}
\end{figure}

The rich phase diagram implies a variety of quantum phase transitions among the phases discussed above. Here we focus on transitions between symmetry-breaking phases and gapped symmetric phases, as well as transitions out of the gapless phases. Among the symmetry-breaking phases, CDW$_2$ and SP both break a \(\mathbb{Z}_2\) symmetry, although in different ways: CDW$_2$ breaks lattice translation symmetry through a period-2 density modulation, whereas SP breaks inversion symmetry through a density imbalance between the two sublattices. Their transitions to neighboring symmetric phases are therefore expected to belong to the Ising universality class. To confirm this, we perform Binder-cumulant data collapse using the corresponding order parameters, as shown in Figs.~\ref{fig:cdwtransitions}(a)--(c). Since the Binder cumulant is dimensionless, curves for different system sizes must cross near the critical point. Along the cut at \(u=9\) with \(t_1=2t_2\) in Fig.~\ref{fig:phasediagram1}(b) , which passes from the trivial phase through SP and into SPT$_1$, we indeed observe two crossings in Fig.~\ref{fig:cdwtransitions}(a), signaling two phase transitions. Optimizing the data collapse gives \(t_{1c}=3.570\) and \(\nu=0.98\) for the trivial--SP transition, and \(t_{1c}=3.670\) and \(\nu=0.96\) for the SP--SPT$_1$ transition. Repeating the same analysis for the SP--SPT$_2$ transition along the cut at \(t_1=2.5\) and \(u=6\) in Fig.~\ref{fig:phasediagram2}(c) gives \(t_{2c}=2.3744\) and \(\nu=0.96\), as shown in Fig.~\ref{fig:cdwtransitions}(b). For the CDW$_2$--SPT$_1$ transition along the cut at \(u=-13\) and \(t_1=2t_2\) in Fig.~\ref{fig:phasediagram1}(b), we obtain \(t_{1c}=3.2121\) and \(\nu=1.01\), shown in Fig.~\ref{fig:cdwtransitions}(c). All of these values are consistent with the Ising exponent \(\nu=1\).

By contrast, CDW$_4$ breaks a \(\mathbb{Z}_4\) translation symmetry, so the transitions out of this phase are generally expected to belong to the Ashkin-Teller (AT) universality class \cite{AshkinStatistics1943PR}. For the CDW$_4$--trivial transition along the cut \(t_2=0.55\), \(u=-4\) in Fig.~\ref{fig:phasediagram2}(a), the optimal collapse shown in Fig.~\ref{fig:cdwtransitions}(d) gives \(t_{1c}=0.45368\) and \(\nu=0.71\). For the CDW$_4$--SPT$_2$ transition along the cut \(t_1=0.1\), \(u=-4\) in the same phase diagram, the best collapse yields \(t_{2c}=0.7600\) and \(\nu=0.97\). Both values fall within the expected AT range \(\nu\in[2/3,1]\) \cite{KohmotoHamiltonian1981PRB}. The latter value is close to the Ising limit, which is natural because for such a small \(t_1\) the odd and even unit cells on each sublattice are nearly decoupled, so the melting of CDW$_4$ toward SPT$_2$ approximately resembles two weakly coupled Ising transitions. We also note that the absence of a direct CDW$_2$--SPT$_2$ transition reflects the incompatibility of their microscopic mechanisms. SPT$_2$ is stabilized by coherent bonding on the \(t_2\) links, whereas CDW$_2$ locks the same links into density configurations such as \(\ket{11}\) or \(\ket{00}\), thereby suppressing charge fluctuations. Recovering bond coherence therefore requires melting the CDW$_2$ order, which naturally leads either to an intervening symmetric regime or to other symmetry-breaking phases favored by \(t_2\) in the presence of interactions, such as CDW$_4$.

To further confirm these universality classes, we analyze the entanglement entropy at the critical points. Figure~\ref{fig:cdwtransitionsvns} shows that \(S_{\rm vN}\) scales linearly with \(\ln L\) for all transition points identified in Fig.~\ref{fig:cdwtransitions}, consistent with CFT predictions. For the four transitions out of the SP and CDW$_2$ phases shown in Figs.~\ref{fig:cdwtransitions}(a)--(c), the extracted central charges are \(c=0.517\), \(0.497\), \(0.545\), and \(0.489\), in good agreement with the Ising value \(c=1/2\). For the two transitions out of CDW$_4$, the extracted central charges are \(0.976\) and \(1.024\), respectively, consistent with the AT value \(c=1\). Taken together, the Binder-cumulant collapses and entanglement-entropy scaling provide strong evidence that the transitions out of CDW$_2$ and SP belong to the Ising universality class, while those out of CDW$_4$ are described by AT criticality.

Finally, we examine the transitions out of the gSPT phase into the trivial and SPT$_2$ phases, which are expected to be of BKT type. Approaching a BKT point from the gapped side, the correlation length diverges through an essential singularity, so the bulk gap should follow the scaling form in Eq.~\eqref{eq:dEuniversalfunction}. To probe this behavior, we calculate the charge-1 gap \(\Delta_1\). On the trivial side, the ground state is nondegenerate, so \(\Delta_1\) can be obtained directly from the ground-state energies in the \(N=L\), \(N=L+1\), and \(N=L-1\) particle-number sectors. On the SPT$_2$ side, however, proximate degenerate boundary modes lead to nearly degenerate states with different boundary configurations when particles are added or removed. To isolate the bulk excitation, we add boundary terms \(-\hat{n}_{A,1}-\hat{n}_{A,2}+\hat{n}_{B,L-1}+\hat{n}_{B,L}\) to energetically separate the nearly degenerate boundary sectors, and then compute \(\Delta_1\) for \(L=256,384,512\). The results along the cut \(t_1=0.5\), \(u=-2\) in Fig.~\ref{fig:phasediagram2}(b) are shown in Figs.~\ref{fig:bkttransitions}(a) and (b). The rescaled gap \(L\Delta_1\) exhibits an extended near-coalescence inside the gSPT phase, bounded by two crossing points for different system sizes, corresponding to transitions into the trivial phase at smaller \(t_2\) and into SPT$_2$ at larger \(t_2\). Data collapse of \(L\Delta_1\) gives the two BKT points at \(t_{2c}=0.3336\) and \(0.4303\). We also extract the Luttinger parameter \(K_{\rm pair}\) from the pair correlation function for \(L=512\) along the same cut. Using the BKT criterion \(K_{{\rm pair},c}=1/2\) \cite{GiamarchiQuantum2003}, we obtain \(t_{2c}=0.3377\) and \(0.4278\), in good agreement with the gap analysis.

For comparison, we further extract the Luttinger parameter from correlation functions of \(L=512\) along the same cut \(t_1=0.5\), \(u=-2\) in the smaller-\(t_2\) regime of Fig.~\ref{fig:phasediagram2}(b), corresponding to the transitions from SPT$_1$ to LL to trivial, and along the cut \(t_1=0.3\), \(u=-4\) in Fig.~\ref{fig:phasediagram2}(a), corresponding to the transitions from CDW$_2$ to pLL to CDW$_4\). Using the critical values \(K_c=2\) for LL and \(K_{{\rm pair},c}=1/2\) for pLL, we obtain \(t_{2c}=0.115\) and \(0.194\) for LL, and \(t_{2c}=0.173\) and \(0.233\) for pLL, both consistent with the positions where \(S_{\rm vN}\) starts to show clear changes. Overall, the agreement between gap-collapse analysis and the critical values of the Luttinger parameters provides strong evidence that the transitions out of LL, pLL, and gSPT are all governed by BKT criticality.

\section{Conclusion} \label{sec:conclusion}

In this work, we have systematically investigated the quantum phases of the half-filled generalized interacting SSH model with intra-cell, NN, and NNN inter-cell hoppings, together with an intra-cell interaction. By combining large-scale DMRG calculations with analyses of entanglement, edge properties, correlation functions, and nonlocal SOPs, we established the global phase diagram and identified a rich set of phases, including two gapped SPT phases, three symmetry-breaking phases, conventional gapless phases with single-particle or two-particle power-law correlations, and a gapless SPT phase.

For the gapped regimes, we showed that the trivial phase, SPT$_1$, and SPT$_2$ can be clearly distinguished by their SOPs, ES, and boundary properties. SPT$_1$ exhibits an exact two-fold ES degeneracy and quantized boundary excess charge \(\pm 1/2\), while SPT$_2$ shows an exact two-fold ES degeneracy together with a proximate four-fold low-lying ES structure. The ES structure further shows that interacting SPT$_2$ is not simply two decoupled copies of SPT$_1$. Its lowest-energy edge sector can have boundary excess charge \(Q=0\) or \(Q=\pm 1\), depending on the inversion properties of the boundary configuration. In the strong-interaction regime, we identified CDW$_2$, CDW$_4$, and SP phases that spontaneously break \(\mathbb{Z}_2\) translation, \(\mathbb{Z}_4\) translation, and inversion symmetry, respectively, and clarified their relation to the neighboring symmetric phases through local order parameters, SOPs, boundary charges, and entanglement diagnostics.

For the gapless regimes, we characterized the conventional LL and pLL phases through their correlation functions, charge gaps, and central charge. In particular, the LL phase shows algebraic single-particle and pair correlations, while the pLL phase has exponentially decaying single-particle correlations, algebraically decaying pair correlations, and a finite single-particle gap. Most importantly, we identified a gSPT phase that combines a \(c=1\) gapless bulk with nontrivial topological boundary structure. This phase is characterized by a finite current SOP, a two-fold ES degeneracy, neutral and Schmidt gaps that decrease exponentially with system size, exponentially decaying single-particle correlations, algebraically decaying pair correlations, and algebraically localized edge current modes. Its properties are consistent with an intrinsically gapless topological phase protected by an anomalous low-energy symmetry realization.

We also determined the universality classes of representative quantum phase transitions in the phase diagram. Transitions out of the CDW$_2$ and SP phases are consistent with Ising criticality, while transitions out of CDW$_4$ are consistent with Ashkin-Teller criticality. The transitions out of the LL, pLL, and gSPT phases are consistent with BKT behavior, as supported by both gap scaling and Luttinger-parameter analysis. These results demonstrate that the generalized interacting SSH model provides a minimal and versatile setting in which gapped and gapless topological phases, symmetry-breaking orders, and multiple types of quantum criticality can coexist and compete.

More broadly, our results provide a useful framework for exploring unconventional interacting topological matter in one dimension and may provide a useful reference for experimental investigations of novel topological phenomena in SSH-type platforms. In particular, recent experiments have realized SSH physics and related topological edge phenomena in synthetic photonic dimensions, Rydberg-atom synthetic dimensions, cold-atom systems, and trapped-ion chains, highlighting the growing accessibility of such models in controllable quantum simulators \cite{LiDirect2023LSA,DeLeseleucObservation2019S,LuWavepacket2024PRA,XieTopological2019nQI,MeierObservation2016NC,LederRealspace2016NC,AtalaDirect2013NP,NevadoTopological2017PRL}.


\begin{acknowledgments}
We thank Jian-Song Pan, Zi-Jian Xiong, and Hai-Yuan Zou for helpful discussions. J.Z. acknowledges support from the National Natural Science Foundation of China under Grant No. 12304172 and from the Chongqing Natural Science Foundation under Grant No. CSTB2024YCJH-KYXM0064. This work was also supported in part by the National Natural Science Foundation of China under Grant No. 12547101.
\end{acknowledgments}

%

\end{document}